\documentstyle[prb,aps,epsf]{revtex}

\newcommand{\cm}{\mbox{cm${}^{-1}$}}

\newcommand{\degree}{${}^{\circ}$}

\begin{document}
%
\twocolumn[ \hsize\textwidth\columnwidth\hsize\csname @twocolumnfalse\endcsname
\title{Extensive infrared spectroscopic study of CuO: signatures of strong spin-phonon
interaction and structural distortion}

\author{A.B.Kuz'menko$^{1}$, D.~van der Marel$^{2}$, P.J.M.~van Bentum$^{3}$,
E.A.Tishchenko$^{1}$, C.Presura$^{2}$ and A.A.Bush$^{4}$}

\address{$^{1}$P.L.Kapitza Institute for Physical Problems RAS,
Kosygina str., 2, Moscow, 117334, Russia}

\address{$^{2}$Solid State Physics Laboratory, University of Groningen, Nijenborgh 4,
9747 AG Groningen, The Netherlands}

\address{$^{3}$High Field Magnet Laboratory, University of Nijmegen, 6525 ED,
Nijmegen, The Netherlands}

\address{$^{4}$Moscow State Institute of Radiotechnics, Electronics and Automation,
Vernadskogo pr.  78, Moscow, 117464, Russia}
\date{10 October 1999}
\maketitle
\begin{abstract}
Optical properties of single-crystal monoclinic CuO in the range 70 - 6000 \cm\
were studied at temperatures from 7 to 300 K. Normal reflection spectra were
obtained from the (001) and (010) crystal faces thus giving for the first
time separate data for the $A_{u}$ and $B_{u}$ phonon modes excited in the purely
transverse way (TO modes). Mode parameters, including polarizations of the $B_{u}$
modes not determined by the crystal symmetry, were extracted by the dispersion
analysis of reflectivity curves as a function of temperature. Spectra of all the
components of the optical conductivity tensor were obtained using the
Kramers-Kronig method recently extended to the case of the low-symmetry crystals. The
number of strong phonon modes is in agreement with the factor-group analysis for
the crystal structure, currently accepted for the CuO. However, several "extra"
modes of minor intensity are detected; some of them are observed in the whole
studied temperature range, while existence of others becomes evident at low
temperatures. Comparison of frequencies of "extra" modes with the available
phonon dispersion curves points to possible "diagonal" doubling of the unit cell
\{{\bf a}, {\bf b}, {\bf c}\} $\rightarrow$ \{{\bf a}+{\bf c}, {\bf b}, {\bf
a}-{\bf c}\} and formation of the superlattice. The previously reported softening
of the $A^{3}_{u}$ mode ($\sim$ 400 \cm) with cooling at $T_{N}$ is found to be
$\sim$ 10 \% for the TO mode. The mode is very broad at high temperatures and
strongly narrows in the AFM phase. We attribute this effect to strong resonance
coupling of this mode to optical or acoustic bi-magnons and reconstruction of the
magnetic excitations spectrum at the N\'eel point. A significant anisotropy of
$\epsilon^{\infty}$ is observed: it was found to be 5.9 along the {\bf b}-axis,
6.2 along the {[}101{]} chains and 7.8 the {[}10$\overline{1}${]} chains. The
"transverse" effective charge is more or less isotropic; its value is about 2
electrons.

\


\

\

\end{abstract}

%

]
 \vspace{2cm}

\narrowtext

\section{Introduction}

Since 1986 the interest to cupric oxide CuO has been mostly governed by its close
relation to the problem of high-$T_{c}$ superconductivity. In addition to the
role of the parent compound of all the high-$T_{c}$ materials with CuO$_{2}$
planes, it has a number of physical and chemical features common to several
undoped antiferromagnetic (AFM) cuprates, (e.g. La$_{2}$CuO$_{4}$,
YBa$_{2}$Cu$_{3}$O$_{6}$): similar copper coordination and electronic state, Cu-O
distances, values of localized magnetic moments, superexchange constants,
low-dimensionality of magnetism etc.

CuO, however, is a quite interesting system in its own right. Although Cu$^{2+}$
ions are expected to be in the $3d^{9}$ state with one 3$d$-hole per atom, this
transition metal (TM) oxide is a strongly correlated insulator of the
"charge-transfer" type according to the theory of Zaanen, Sawatzky and Allen
\cite{ZSA}; the holes are well localized forming local magnetic moments. CuO
undergoes a 2-stage magnetic transition: at $T_{N1}$ = 230 K an incommensurate
magnetic structure is observed, while at $T_{N2}$ = 213 K magnetic moments order
parallel to the {\bf b}-axis antiferromagnetically along the
${[}10\overline{1}{]}$ chains and ferromagnetically along the [101] chains
\cite{Yang}. From the analysis of the spin-wave velocity \cite{Yang,Ain} it was
found that the exchange constant along the ${[}10\overline{1}{]}$ chains (60 - 80
meV) is several times larger than this value along any other direction. The
anomalous temperature dependence of the magnetic susceptibility \cite{OKeeffe}
points to low-dimensional, or, at least, highly anisotropic character of magnetic
interactions and persistency of spin correlations at temperatures well above the
N\'eel point \cite{Kobler,Chattopadhyay}.

Another feature of the cupric oxide is the low-symmetry monoclinic lattice, which
distinguishes it from the other TM monoxides, e.g. MnO, FeO, CoO and NiO with the
rock-salt structure. It is a prominent manifestation of the Jahn-Teller effect:
in the high-symmetry octahedral position characteristic to the cubic structure,
the Cu$^{2+}$ ion would have degenerate $d_{x^{2}-y^{2}}$ and $d_{z^{2}}$
orbitals, which is energetically unfavourable, and therefore tend to displace
away from the symmetry position. This tendency is so strong that CuO has not just
a distorted cubic lattice, but a completely different monoclinic tenorite
structure.

Several groups \cite{Popovic,Hanuza,Degiorgi,Kliche,Guha,Narang,Homes} have
reported results of infrared (IR) spectroscopic studies of powder as well as
single-crystal specimens of CuO. The interpretation of infrared spectra was
always embarrassed by the low crystal symmetry, especially for the case of
polycrystalline samples. Kliche and Popovic \cite{Kliche} have measured infrared
spectra of sintered powder samples as a function of temperature and for the first
time assigned strong IR-active modes to the species $A_{u}$ and $B_{u}$ by
comparison of frequencies with those in PdO. They also reported an additional
broad mode at about 414 \cm\ the intensity of which increases drastically with
cooling down below $T_{N}$ and suggested that it is a zone-boundary phonon mode
which becomes IR-active because of IR absorption from AFM superstructure. It is
evident now that it was a manifestation of the anomalous softening of the
$A_{u}^{3}$ mode reported by Homes {\em et al} \cite{Homes}.

So far it was a serious problem to obtain single crystals of CuO suitable for
quantitative infrared studies. Guha {\it et al} \cite{Guha} has succeeded to
measure infrared polarized spectra of single crystals of CuO at room temperature
and account for low-symmetry effects in data analysis. They have measured
reflectivity from the (1$\overline{1}$0) natural face and modelled spectra by
the dielectric function formulas adapted to monoclinic crystals \cite{Belousov}.
However, due to inconvenient crystal orientation in their experiment mixed LO-TO
modes were excited, the properties of which depend on the wave vector direction.

Homes {\it et al} \cite{Homes} were the first to present single-crystal infrared
spectra as a function of temperature. Again, however, only the
(1$\overline{1}$0) crystal surface was accessible for optical experiments,
and mixed LO-TO modes were actually observed. An appreciable (about 5 \%) sharp
softening of the $\sim$ 440 \cm\ reststrahlen band for the {\bf E} $\perp$ {\bf
b} upon cooling down was definetely registered at the N\'eel transition. Spectra were
fitted with introduction of $3\,A_{u}$ and $3\,B_{u}$ phonon
modes only. No new phonon structures at the magnetically ordered phase were
reported indicating absence of a crystal superlattice below $T_{N2}$.

This statement sounds puzzling in a view of observation by Chen {\it et al}
\cite{Chen}\ of five new modes at low temperatures in the Raman spectra. Authors
have assigned these modes to folded phonons; as a folding mechanism, a strong
spin-phonon interaction was proposed. The most intense new mode 240 \cm\ hardens
strongly at cooling down, which was attributed \cite{Chen}\ to an additional
lattice rigidity due to magnetization.

There is a serious inconsistency concerning structure and parameters of IR-active
phonon modes, especially at high frequencies. For instance, the deviations in
resonance frequency of these modes reported by different groups are too
significant to be explained by experimental errors, isotope effect, crystal
non-stoiochiometry etc. In our opinion, the explanation lies in the intermediate
LO-TO nature of the observed modes and corresponding uncertainty of phonon
parameters, especially for the high-frequency intense modes with large LO-TO
splitting. Moreover, no infrared data so far were reported where the $A_{u}$ and
$B_{u}$ modes were completely separated. Parameters of the $A_{u}$ modes were
extracted at best from the single-crystal spectra for ${\bf E}\perp c$ where
the $B_{u}$ modes are also present.

In this paper we aimed to resolve this uncertainty by separate measurement of the
characteristics of purely TO $A_{u}$ and $B_{u}$ modes. For monoclinic crystals
the only option for observation of the $B_{u}$ TO modes is to measure normal
reflectivity from the (010) face (the {\bf ac}-plane). To study the $A_{u}$ TO modes any
crystal plane containing the {\bf b}-axis, e.g.  (001) face, may suffice. We
succeeded to obtain these crystal faces with a sufficiently large area, allowing
to perform reliable measurements and quantitative analysis of the data as
described below.

\section{Crystal structure and factor-group analysis}

\begin{figure}[t]

\centerline{\epsffile{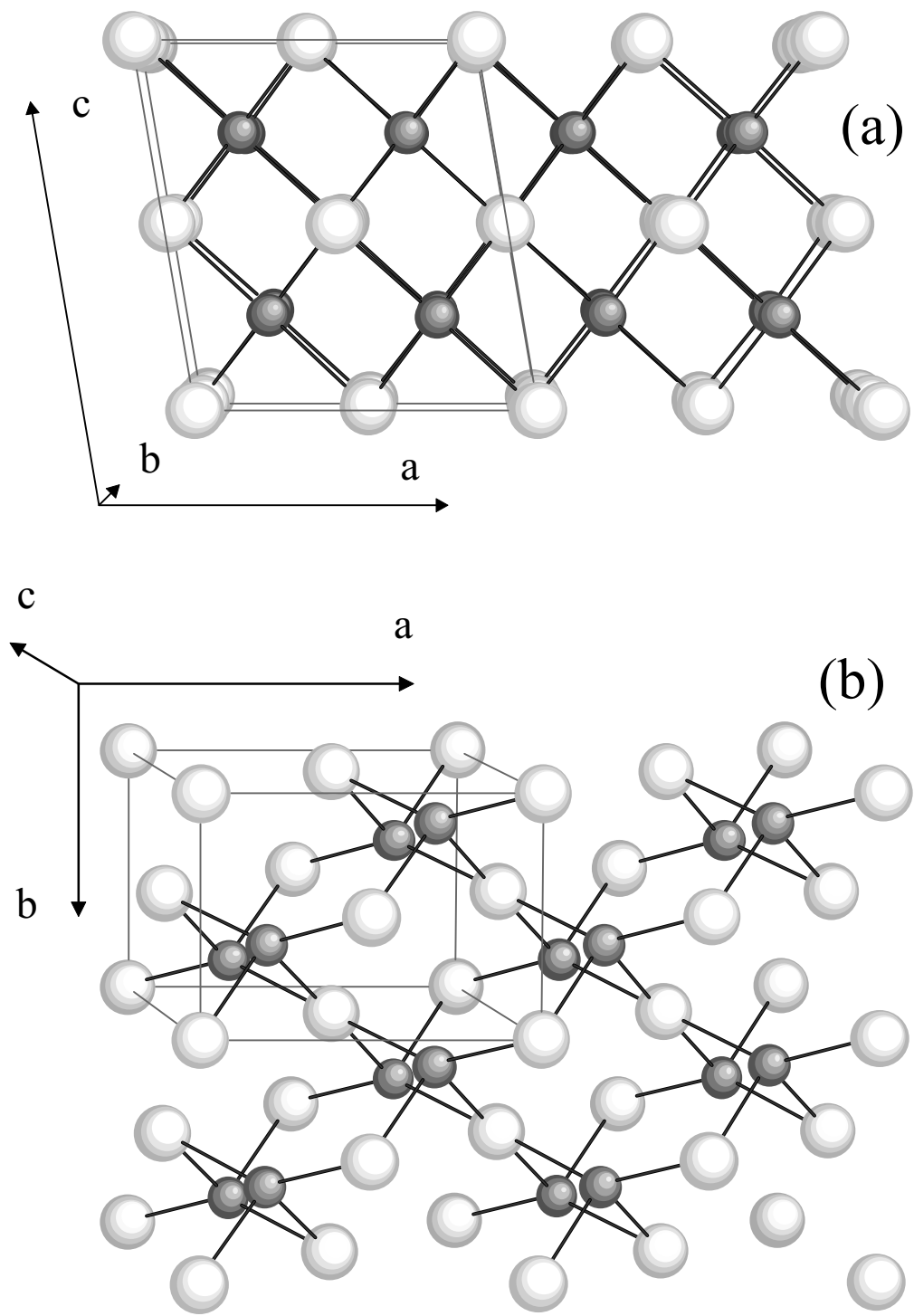}}

\caption{The crystal structure of CuO (after Asbrink and Norrby). Different
projection views of the $2{\bf a} \times {\bf 2b} \times {\bf c}$ cell array are
shown: the {\bf ac}-plane view (a) and the {\bf ab}-plane view (b). The unit cell
containing four CuO units is indicated by the parallelepiped. The copper atoms
are light gray, the oxygen atoms are dark gray.}

\label{FigStruct}
\end{figure}

\begin{figure}[t]

\centerline{\epsffile{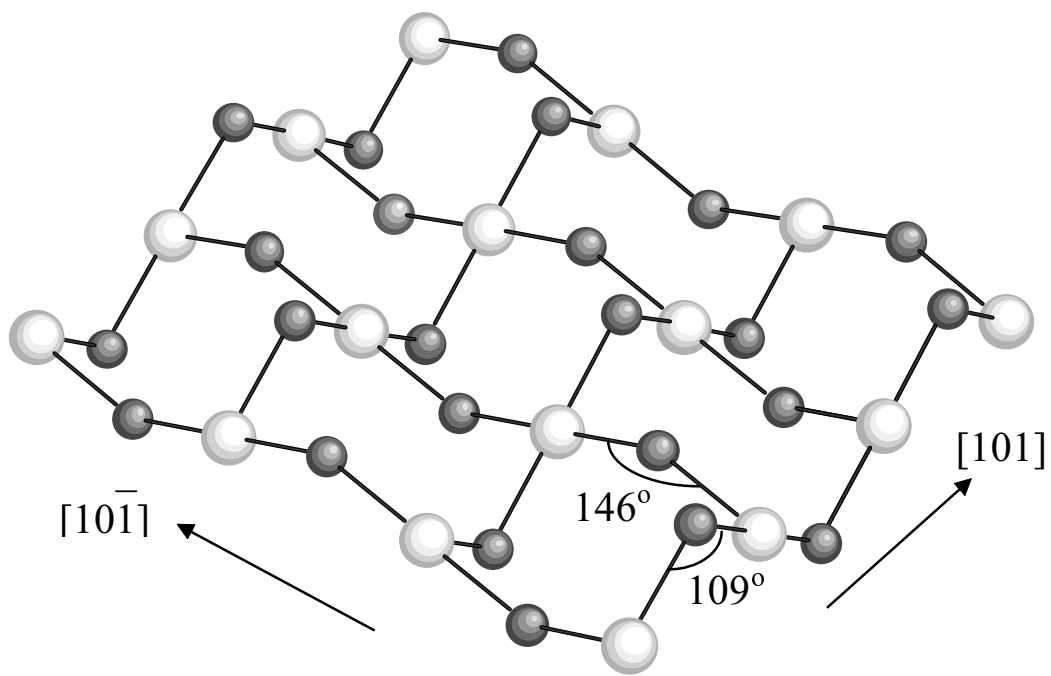}}

\caption{The Cu-O chains running along the {[}101{]} and the
{[}10$\overline{1}${]} directions. The copper and oxygen atoms are light gray and
dark gray correspondingly. The Cu atoms with the same {\bf b}-coordinates only
are shown.}

\label{FigChains}
\end{figure}

Cupric oxide CuO, unlike other TM monoxides, crystallizes in a low-symmetry
monoclinic tenorite structure (Fig.~\ref{FigStruct}). It is generally accepted,
following {\AA}sbrink and Norrby \cite{AsbrinkNorrby}, that at room temperature
(RT) the space group is C$_{2h}^{6}$ (C2/c); there are four CuO molecules in the
unit cell with dimensions {\bf a} = 4.6837 \AA, {\bf b} = 3.4226 \AA, {\bf c} =
5.1288 \AA, $\beta$ = 99.54\degree\ and two CuO units in the primitive cell; the
copper and oxygen occupy the C$_{i}$ and C$_{2}$ symmetry positions
correspondingly. Each copper atom is situated in the center of the oxygen
parallelogram. Each oxygen atom, in turn, has a distorted tetrahedral copper
coordination. The adjacent CuO$_{4}$ parallelograms form two sets of ribbons
propagating along the {[}110{]} and the {[}1$\overline{1}$0{]} directions. The
structure can be also considered as being composed from two types of zig-zag Cu-O
chains running along the {[}101{]} and the {[}10$\overline{1}${]} directions
(Fig.~\ref{FigChains}). The Cu-O-Cu angle is 146\degree\ in the
{[}10$\overline{1}${]} chains and 109\degree\ in the {[}101{]} chains.

For the C$_{2h}^{6}$ space group the factor-group (FG) analysis \cite{Rousseau} gives the
following set of the zone-center lattice modes: $\Gamma =
A_{g}+2B_{g}+3A_{u}+3B_{u}+3\mbox{ translational}$. Out of these, 3 modes
($A_{g}+2B_{g}$) are Raman-active, 6 modes (3$A_{u}+3 B_{u}$) are IR-active. The
$A_{u}$ modes are polarized along the {\bf b}-axis. The dipole moments of the
$B_{u}$ modes lie within the {\bf ac}-plane, but due to the low symmetry their
directions are not exactly determined by the crystal structure.

More recently {\AA}sbrink and Waskowska\cite{AsbrinkWaskowska} have refined the
CuO structure at 196 K and 300 K using the so called "less significant
reflections" in the X-ray data analysis and found that less symmetric space-group
C$_{s}^{4}$ (Cc) is also consistent with the X-ray diffraction data for both
phases. They suggested that the C$_{2h}^{6}$ space-group might result from the
time-averaging or site-averaging of non-equivalent (due to valence fluctuations)
atom positions of lower symmetry. Some lattice distortions, especially changes of
the Cu-O distances, were clearly detected when passing from RT to 196 K. In
general, one can state, that the C$_{2h}^{6}$ space-group is a good approximation
to the real structure of the cupric oxide, but some minor deviations from this do
not contradict to the X-ray data.

\section{Experimental}

\subsection{Sample preparation and characterization}

Single crystals of CuO were obtained from a CuO - PbO - Bi$_{2}$O$_{3}$ melt. The
details are described elsewhere \cite{Bush}. After cooling down the crucible
contained randomly oriented large single-crystal pieces of CuO along with
inclusions of other phases. From this conglomerate the largest CuO single
crystals were extracted and oriented using the X-ray diffraction. As usual,
natural crystal faces were presumably of (110) and (1$\overline{1}$0)
orientation. This face orientation was used in previous papers where infrared
reflectivity measurements were performed. However, for the reasons mentioned
above, we aimed to obtain large enough the (001) (the {\bf ab}-plane) and the
(010) (the {\bf ac}-plane) crystal faces. These two mutually perpendicular faces
were cut on one selected single-crystal sample, which was used for measurement of
all the reflectivity spectra presented in this paper. Cuts were polished with a
fine 0.06 $\mu$m Al$_{2}$O$_{3}$ powder. Microscopic analysis of the surface
\cite{Bush} has shown that the crystal is twinned. Fortunately, one twin
orientation was almost completely dominating; the domains of the alternative twin
orientation form narrow stripes covering less than 5 \% of surface area. Such
domination was also confirmed by the X-ray Laue snapshots, where no detectable
reflections corresponding to the alternative twin orientation were observed. The
structure of twins is such \cite{Kryukova} that the {\bf b}-axis direction is the
same for different twin orientations; all twin reflections are within the {\bf ac}-plane.
So a minor (less than 5 \%) contribution of other twin domains is
possible for the (010) face reflectivity spectra. For the case of reflection
from the (001) face, when {\bf E} $\parallel$ {\bf b}, all twins contribute
in the same way and twinning has no effect. The Lauegram has shown that
(001) and (010) crystal faces were cut with accuracy of 1.8\degree\ and 2.1\degree
respectively. The electron Auger microscopy has shown the presence of only copper
and oxygen atoms on the both crystal faces. The area of the (001) face
suitable for quantitative optical measurements (i.e. containing no impurity
inclusions, having the lowest fraction of the alternative twin orientation), was
$\sim$ 3 mm$^2$; that of the (010) face was $\sim$ 4 mm$^2$.

\subsection{Reflectance measurement}

Infrared reflectivity spectra were measured from 70 to 6000 \cm\ using a Bruker
IFS 113v FT-IR spectrometer. The average angle of incidence was about 11\degree.
A set of different light sources, beamsplitters, polarizers and detectors were
used to cover this frequency range. The mid-infrared (MIR) spectra from 400 to 6000
\cm\ were measured using a globar source, KBr beamsplitter, KRS-5 polarizer and
DTGS and MCT detectors. The far-infrared (FIR) region 70 - 700 \cm\ was studied
with the aid of the Hg lamp, a set of mylar beamsplitters, a polyethylene polarizer
and the helium-cooled Si bolometer.

The polarizer was mounted in the optical path of the incident beam; no additional
polarizers (analyzers) were put inbetween sample and detector. The transmission
properties of the polarizers were measured independently and, when necessary,
special care of the correction for the unwanted polarization leakage was taken.
Polarizer rotation was performed using a computer-controlled mechanical rotator.

An original "three-polarization" measurement technique \cite{Kuzmenko1} was used
involving the measurement of three reflectivity spectra per crystal face for
different polarizations of almost normally incident light: vertical (0\degree),
horizontal (90\degree) and diagonal (45\degree). In principle, the knowledge of
these spectra should be enough to calculate the reflectivity for any other
polarization direction. In particular, the relation
$R(0^{\circ})$+$R(90^{\circ})$ = $R(45^{\circ})$+$R(-45^{\circ})$ should work. We
especially checked the validity of this relation and experimentally proved that
it holds with a good accuracy.

The sample was mounted with a good thermal contact in a continuous-flow cryostat
(Oxford Instruments) with an automatic temperature control. Spectra were measured
at temperatures 300, 250, 240, 230, 220, 210, 200, 180, 150, 100 and 7 K,
so that, special attention has been paid to the range in the vicinity of
$T_{N_{1}}$ = 230 K and $T_{N_{2}}$ = 213 K. The temperature setting accuracy was
about 1 K.

A reference for the absolute reflectivity was provided by {\em in situ}
evaporation of a gold layer on the surface and consecutive repetition of the same
set of measurements for every temperature. Such a procedure has compensated
errors associated with not only non-ideality of the sample face but also the
thermal deformation of the cryostat cold finger. To account for a possible drift
of a single-beam intensity due to source and detector instability, every
sample-channel measurement was accompanied by a measurement of the intensity of
the light beam passed via the second channel without reflection from the sample.

\section{Spectra treatment}

\subsection{The dispersion analysis}

Let us introduce the orthogonal system of coordinates \{${\bf xyz}$\}: ${\bf x} \parallel
{\bf a}$, ${\bf y} \parallel {\bf b}$, ${\bf z} \perp {\bf a}$, ${\bf z} \perp {\bf b}$ so that there
is a slight inclination ($\sim$ 9.5\degree) between axes {\bf z} and {\bf c}. Due to
the monoclinic symmetry the whole 3D dielectric tensor $\hat{\epsilon}$ is the
composition of two components: the scalar $\epsilon_{b}=\epsilon_{yy}$ along the
{\bf b}-axis and the 2D tensor $\hat{\epsilon}_{ac}=\left[\begin{array}{cc}
\epsilon_{xx} & \epsilon_{xz}
\\ \epsilon_{zx} & \epsilon_{zz}
\end{array}\right]
$ within the {\bf ac}-plane ($\epsilon_{xz}=\epsilon_{zx}$ is expected without external magnetic field).
The dispersion formulas are:

\begin{equation}\label{epsb}
\epsilon_{b}(\omega)=\epsilon_{b}^{\infty}+\sum_{i, A_{u}}
\frac{\omega_{\mbox{\scriptsize p},i}^{2}}{\omega_{\mbox{\scriptsize TO},i}^{2}-\omega^{2}-\mbox{i}\gamma_{i}\omega}\mbox{
, }
\end{equation}

\begin{eqnarray}\label{epsac}
\hspace{10mm}\hat{\epsilon}_{ac}(\omega)&=&\hat{\epsilon}_{ac}^{\infty} +
\sum_{i, B_{u}}
\frac{\omega_{\mbox{\scriptsize p},i}^{2}}{\omega_{\mbox{\scriptsize TO},i}^{2}-\omega^{2}-\mbox{i}\gamma_{i}\omega}\times
\nonumber \\ &\times& \left[
\begin{array}{cc}
\cos^2 \theta_{i} & \cos \theta_{i} \sin \theta_{i} \\
\cos\theta_{i}\sin\theta_{i} & \sin^2 \theta_{i} \\
\end{array}
\right]\mbox{,}
\end{eqnarray}

\noindent where $\omega_{\mbox{\scriptsize TO},i}$ - the transverse frequency,
$\omega_{\mbox{\scriptsize p},i}$ - the plasma frequency, $\gamma_{i}$ - the
linewidth of the $i$-th mode, $\theta_{i}$ - the angle between the dipole moment
of the $i$-th mode and the $x$-axis (for the $B_{u}$ modes only), $\epsilon_{b}^{\infty}$ and
$\hat{\epsilon}_{ac}^{\infty}$ are the high-frequency dielectric tensors. The
{\bf b}-axis complex reflectivity $r_{b}$ and the reflectance of the (001)
plane for {\bf E} $\parallel$ {\bf b} are expressed via the
dielectric function:

\begin{equation}\label{rb}
r_{b}(\omega)=\frac{1-\sqrt{\epsilon_{b}(\omega)}}{1+\sqrt{\epsilon_{b}(\omega)}}\mbox{
,\hspace{5mm}} R_{b}(\omega)=|r_{b}(\omega)|^{2}\mbox{.}
\end{equation}

The complex reflectivity tensor $\hat{r}_{ac}$ can be expressed via the
dielectric tensor $\hat{\epsilon}_{ac}$ by the matrix formula, which is formally
analogous to (\ref{rb}):

\begin{equation}\label{rac}
\hat{r}_{ac}(\omega)=(\hat{1}-\sqrt{\hat{\epsilon}_{ac}(\omega)})\cdot
(\hat{1}+\sqrt{\hat{\epsilon}_{ac}(\omega)})^{-1}\mbox{,}
\end{equation}

\noindent where $\hat{1}$ is the unity tensor. The matrix square root naturally
means, that the matrix is first reduced to the diagonal form by a proper
rotation, the square root is then taken from each diagonal element, and finally
it is rotated back to the initial coordinate system. The "-1" exponent implies
calculation of the inverse matrix.

The reflectance of the (010) plane depends on the direction of the incident
light polarization ${\bf e}={\bf E}/|{\bf E}|$:

\begin{equation}\label{R_ac}
R_{ac}(\omega, {\bf e})=|\hat{r}_{ac}(\omega)\,{\bf e}|^{2}\mbox{.}
\end{equation}

\noindent The reflectances for values 0\degree, 45\degree and 90\degree\ of the
angle between the electric field vector and the {\it x}-axis used in the
"three-polarization" measurement scheme\cite{Kuzmenko1} are expressed in terms of
the components of $\hat{r}_{ac}$:

\begin{eqnarray}\label{R004590}
R_{00}(\omega)&=&|r_{xx}(\omega)|^{2} + |r_{xz}(\omega)|^{2} \nonumber
\\ R_{45}(\omega)&=&(|r_{xx}(\omega)+ r_{xz}(\omega)|^{2} +\nonumber
\\ &+&|r_{xz}(\omega)+ r_{zz}(\omega)|^{2})/2 \nonumber \\
R_{90}(\omega)&=&|r_{xz}(\omega)|^{2} + |r_{zz}(\omega)|^{2}\mbox{.}
\end{eqnarray}

The phonon parameters can be obtained by fitting of the reflectance spectra using
the written above formulas. To obtain the $A_{u}$ modes parameters the
$R_{b}(\omega)$ spectrum is fitted. The characteristics of the $B_{u}$ modes,
including unknown angles $\theta_{i}$, can be extracted by simultaneous fitting
of three spectra $R_{00}(\omega)$, $R_{45}(\omega)$, $R_{90}(\omega)$ from the
(010) plane.

\subsection{The Kramers-Kronig analysis}

For the case {\bf E} $\parallel$ {\bf b} the Kramers-Kronig (KK) analysis
can be performed in a usual way because one of the dielectric axes is parallel to
this direction. Due to the low crystal symmetry, the directions of the two other
principal dielectric axes in the {\bf ac}-plane depend on the frequency, which
precludes the straightforward application of the KK method to the {\bf ac}-plane
reflectance data. For this case we used a modified version of this technique,
which allows to determine frequency dependence of all the components of the
complex reflectivity tensor $\hat{r}_{ac}$, provided that three reflectance
spectra $R_{00}(\omega)$, $R_{45}(\omega)$ and $R_{90}(\omega)$
are measured in wide enough frequency range. The details of
this KK method generalization are described in Ref.\cite{Kuzmenko2}.

For a correct implementation of the KK integration, the reflectivity in the range
6000 - 37000 \cm\ was measured using the Woollam (VASE) ellipsometer system. At
higher frequencies the $\omega^{-4}$ asymptotics was assumed, as usual. At low
frequencies the reflectivity was extrapolated by a constant value.

\section{Results and analysis}

\subsection{{\bf E} $\parallel$ {\bf b}}

\begin{figure}[t]
\centerline{\epsffile{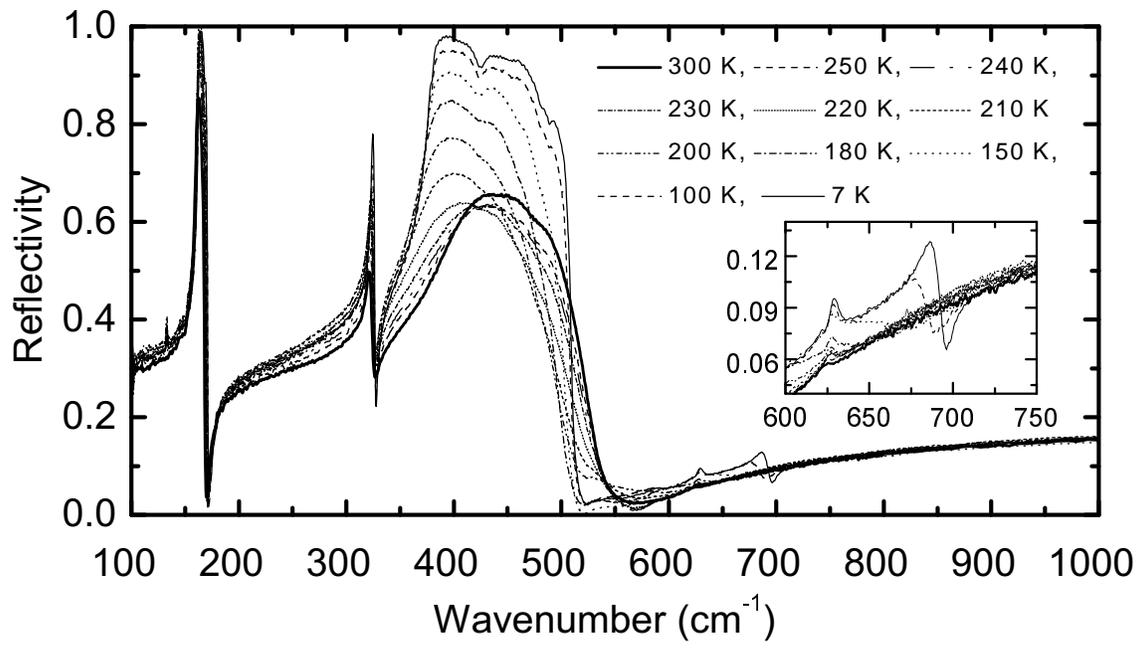}}

\caption{Reflectance spectra for the (001) plane for polarization {\bf E}
$\parallel$ {\bf b} as a function of temperature.}

\label{FigRb}
\end{figure}

The reflectance spectra for the (001) plane, when {\bf E} $\parallel$ {\bf
b}, are shown at Fig.~\ref{FigRb}. In this configuration only the $A_{u}$ TO
modes should be active. Exactly three strong modes are observed: $A_{u}^{1}$
($\sim$ 160 \cm), $A_{u}^{2}$ ($\sim$ 320 \cm) and $A_{u}^{3}$ ($\sim$ 400 \cm),
which confirms the FG-analysis predictions for the established for CuO crystal structure.
The most drastic temperature changes take place in the range 350 - 550 \cm, where the
reststrahlen band corresponding to very intense lattice mode $A_{u}^{3}$ is
situated. The reflectivity maximum elevates from 65 \% to almost 100 \%; its
gravity center moves to lower frequencies upon cooling down the sample. It
indicates, that this mode experiences strong softening and narrowing as a
temperature is decreased.

\begin{figure}[t]
\centerline{\epsffile{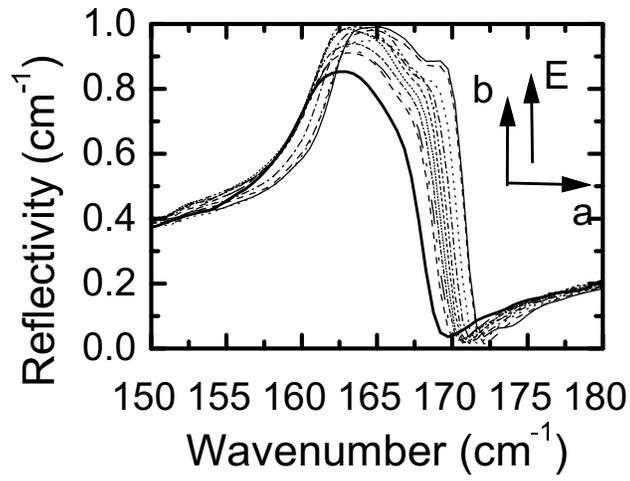}}

\caption{Reflectance spectra for the (001) plane for polarization {\bf E}
$\parallel$ {\bf b} as a function of temperature (enlarged view of the $A_{u}^{1}$ mode region.}

\label{FigMode162}
\end{figure}

In addition to three strong modes, at least five "extra" structures in these
reflectivity spectra are detectable "by eye" (Fig.~\ref{FigRb}). The first
structure is a dip at $\sim$ 425 \cm\ just on the top of the reststrahlen band,
which becomes visually evident below 210 K. The second is a $\sim$ 485 \cm\
structure also on the top of the same reststrahlen band seen at 100 K and 7 K.
The third feature is a $\sim$ 630 \cm\ peak (see inset), which is very small (but
observable) at 300 K, and becomes significant at low temperatures. The
fourth structure is high-frequency mode $\sim$ 690 \cm\ (see inset), which is
very obvious at 7 K (690 \cm) and 100 K (680 \cm), still detectable at 150 K
($\sim$ 650 \cm) and not seen at higher temperatures, probably, because of strong
broadening. The fifth structure is seen at 165 - 170 \cm\ at the top of the
reststrahlen band of the $A_{u}^{3}$ mode (Fig.\,\ref{FigMode162}(a)).
It is better observable at low temperatures; but even at room temperature the
form of the reststrahlen band differs from a single-mode shape.

\begin{figure}[t]
\centerline{\epsffile{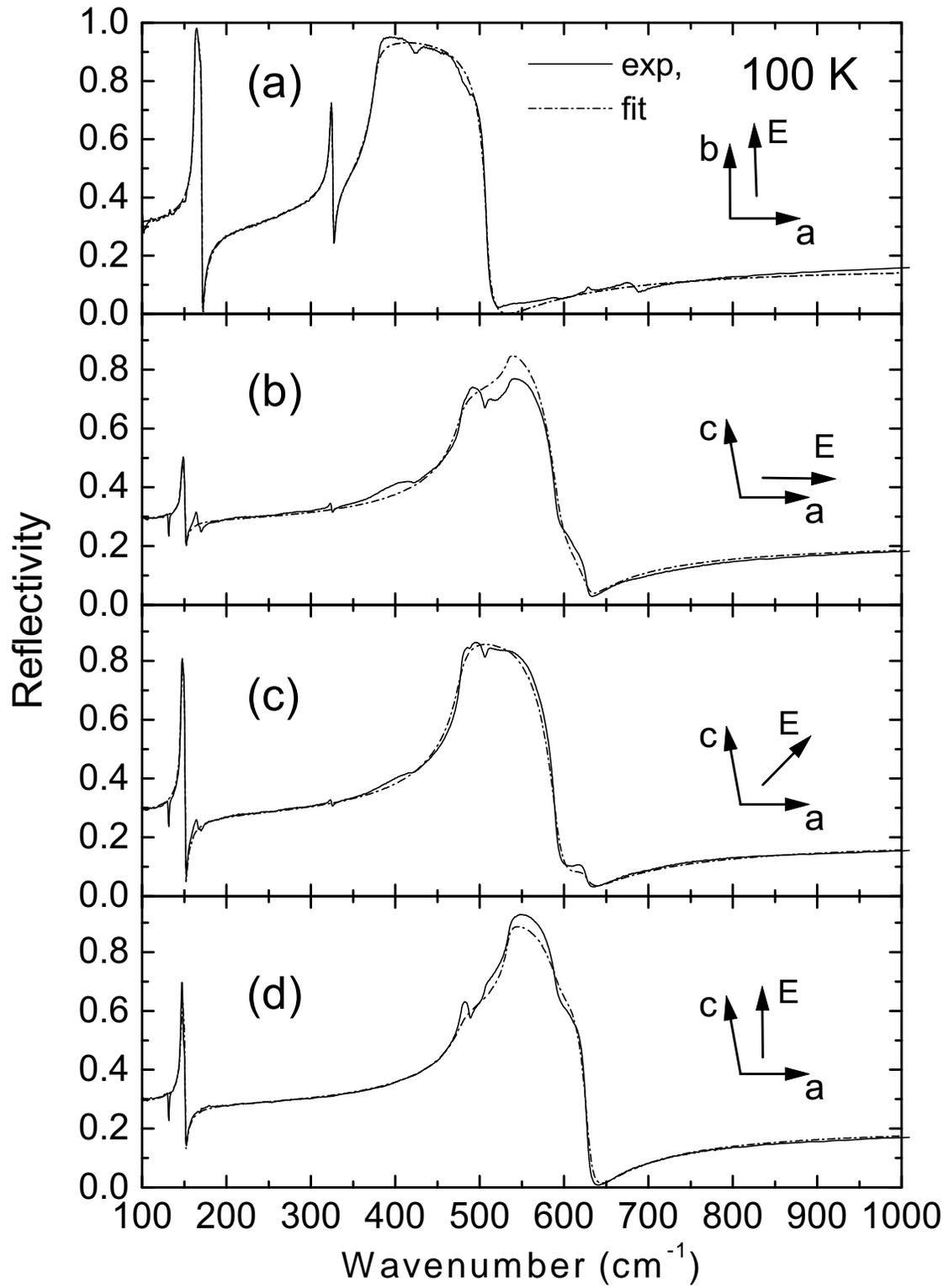}}

\caption{Fitting of the measured spectra for different polarizations at 100 K by
the dispersion formulas.}

\label{FigRfit}
\end{figure}

\begin{figure}[t]
\centerline{\epsffile{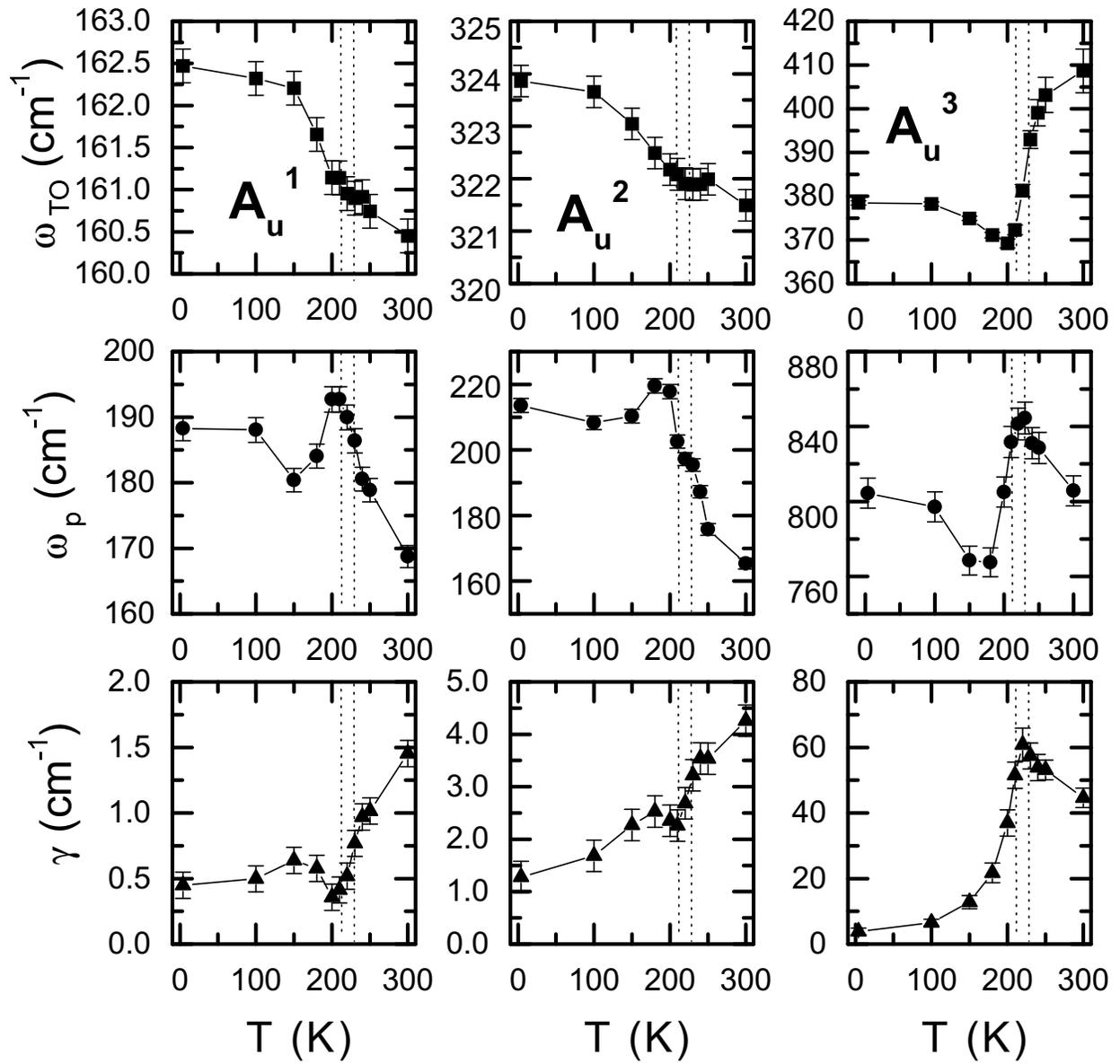}}

\caption{Parameters of the principal $A_{u}$ TO modes, obtained by the 3-mode dispersion analysis
of the reflectance spectra for the (001)-plane, {\bf E} $\parallel$ {\bf b}.
Vertical dotted lines denote $T_{N1}$ and $T_{N2}$.}

\label{FigAuto}
\end{figure}

The observations "by eye" should be accompanied by numerical analysis. The
dispersion analysis of spectra was performed in two stages. On the first stage,
in order to determine the characteristics of the principal modes, we have fitted
the reflectivity curves with introduction of 3 modes only. The experimental
and fitting curves are compared at Fig.~\ref{FigRfit}(a) for the $T$ = 100 K. The
fit quality is good enough; the deviations are observed only in the range of
additional modes. A relative weakness of additional structures ensures small
errors in determination of main mode parameters. The parameters of 3 principal
$A_{u}$ phonon modes obtained by such a fitting as a function of temperature are
shown at Fig.~\ref{FigAuto}.

From Fig.\,\ref{FigAuto} a conclusion can be drawn immediately that the
$A_{u}^{1}$ and $A_{u}^{2}$ modes behave in a quite similar way, while the
highest-frequency $A_{u}^{3}$ mode is significantly different. The $A_{u}^{1}$
and $A_{u}^{2}$ modes are steadily hardening with cooling ($\sim$ 1 \%), with some
increasing of the slope $\partial \omega_{\mbox{\scriptsize TO}}/\partial T$
below $T_{N}$. In contrast, the $A_{u}^{3}$ mode slightly softens with cooling
down to the $T_{N1}$, then undergoes a drastic sharp softening ($\sim$ 10 \%) in
the vicinity of the transition temperature and then hardens with further
cooling from $T_{N2}$ down to the helium temperature. The $A_{u}^{2}$ and
$A_{u}^{3}$ modes are relatively narrow at 300 K and exhibit further narrowing
with cooling with some dip at the transition temperature. On the contrary, the
$A_{u}^{3}$ mode is very broad at 300 K, and broadens more with approaching the
$T_{N}$. Its linewidth has a pronounced maximum inbetween $T_{N1}$ and $T_{N2}$.
In the AFM phase it quickly narrows with cooling down.

\begin{figure}[t]
\centerline{\epsffile{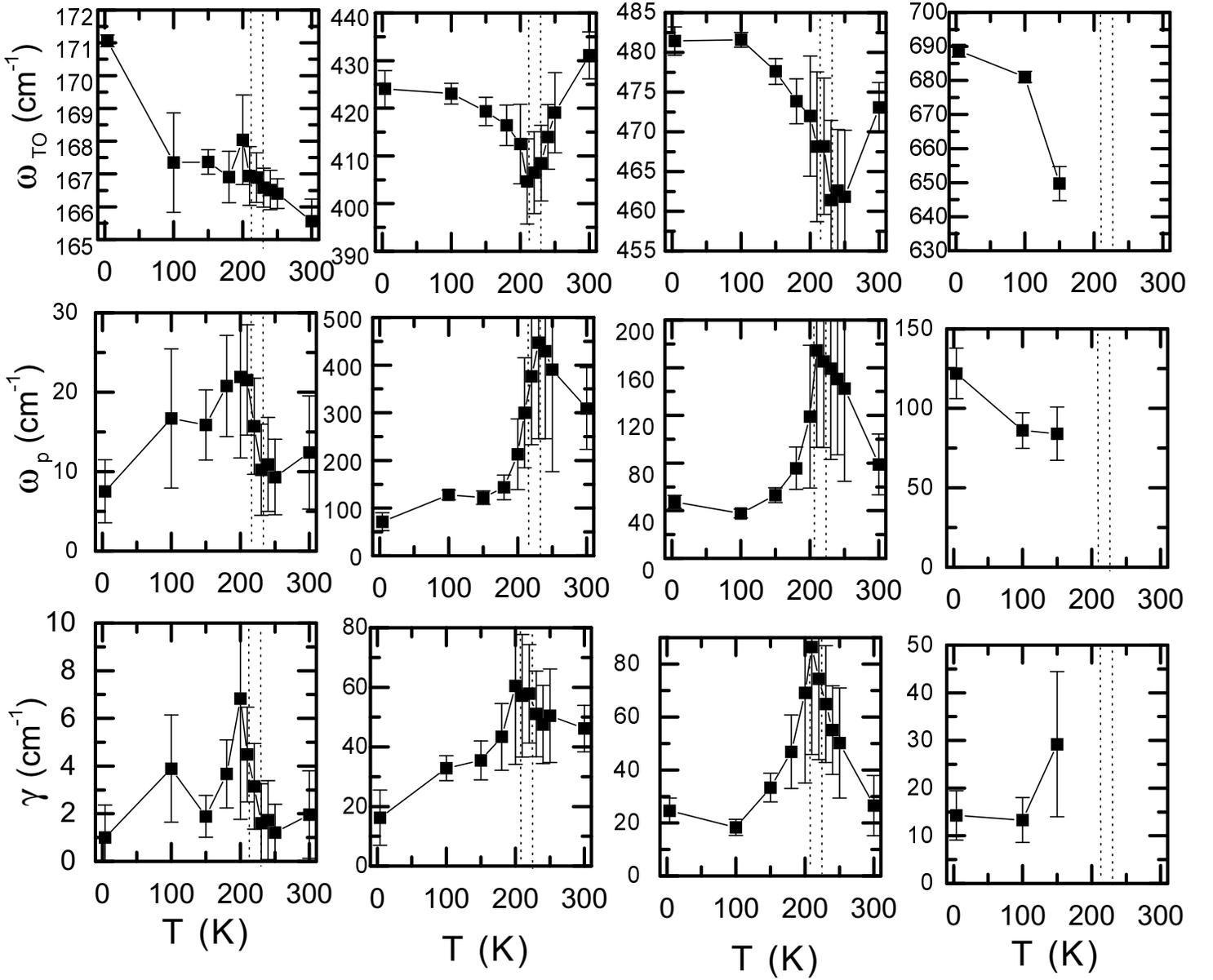}}

\caption{Parameters of the "extra" modes, obtained by the full (both principal and extra modes are
considered) dispersion analysis of the reflectance spectra for the (001)-plane,
{\bf E} $\parallel$ {\bf b}. Error bars indicate parameter confidence limits and reflect possible
correlation of different parameters. Vertical dotted lines denote $T_{N1}$ and $T_{N2}$.}

\label{FigAutoext}
\end{figure}

On the second stage, in order to determine or, at least, estimate parameters of the
mentioned "extra" modes, additional fitting of
spectra with introduction of both principal and "extra" modes has been performed
(see Fig.\,\ref{FigAutoext}). It was possible to fit the structures at 167 \cm,
425 \cm, 485 \cm\ and 690 \cm. The 630 \cm\ peak cannot be fitted by the usual
lorentian term. The first fit was performed at 7 K, when additional structures
are mostly sharp. Other spectra were fitted in sequence 100 K, 150 K, ..., 300 K.
At each step the oscillator parameters, corresponding to the previous temperature
served as initial approximation for the least-squares fitting. Parameter
confidence limits were always calculated by the "covariant matrix" method (see
error bars at Fig.\,\ref{FigAutoext}), which takes into account possible correlation of parameters.
In this way it was possible to extend curves of modes 167 \cm, 425 \cm\ and 480 \cm\ up to room
temperature; the 690 \cm\ mode was fitted only for $T \leq $ 150 K, because at higher temperatures
fitting could not give reasonable values of parameters for this mode.
The errors in determination of additional modes are relatively larger than those
for the principal ones, which is natural, in view of their small intensity. In
general, errors are smaller at lower temperatures.

The temperature dependence of TO frequency additional mode 167 \cm\
is more or less typical for phonons (the value of 171 \cm\ at 7 K is
an artefact of the dispersion analysis with no physical
meaning). On the contrary, the 425 \cm\ and 485 \cm\ modes demonstrate
a puzzling temperature dependence, similar to that of the
$A_{u}^{3}$ mode 400 \cm. First of all, there is no indication of
disappearance of these modes above the AFM transition, although they are not clearly seen "by
eye". Nevertheless, the transition strongly affects parameter values of these modes.
Both modes strongly soften above $T_{N}$ and strongly harden below $T_{N}$. The modes are
narrowing below $T_{N}$ (which facilitates their visual observation),
and are very broad at higher temperatures with a strong maximum at the
transition point. One can content only by a qualitative conclusions,
because the temperature dependencies of these modes are masked by large
error bars. It can be explained by some correlation between
parameters of these modes with ones of the $A_{u}^{3}$ mode. It is
a typical problem for several broad closely located modes.
Therefore, it is unreasonable to attribute physical meaning to increasing of the
plasma frequency of these modes above 200 K: their plasma frequencies
are just subtracted from the plasma frequency of the $A_{u}^{3}$ mode
without significant influence on the fit quality. For the 690 \cm\
mode the dispersion analysis results confirm visual observations:
anomalously strong hardening at cooling and strong broadening with
heating. The latter, probably, precludes satisfactory fitting of this
mode at higher temperatures.

\begin{figure}[t]
\centerline{\epsffile{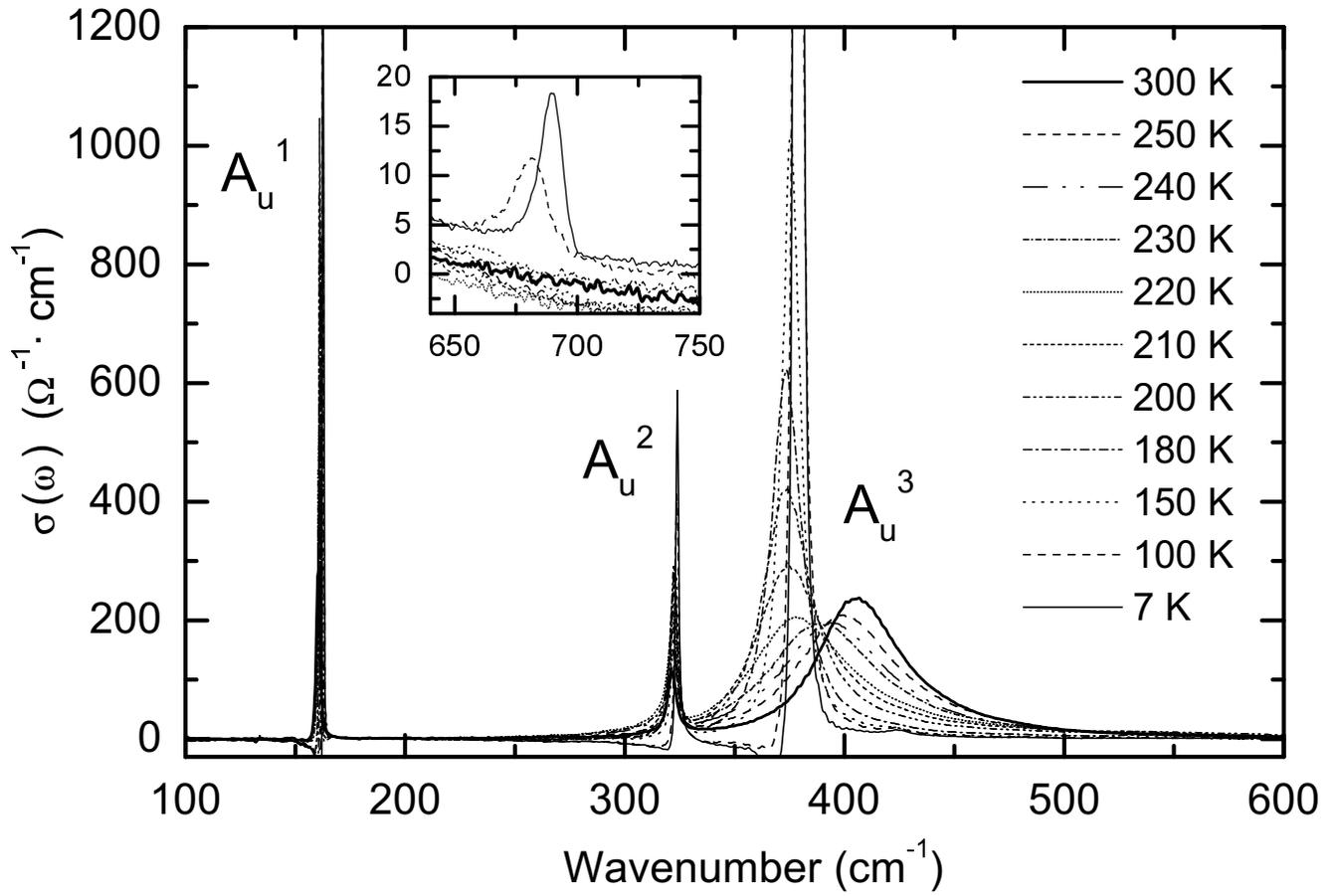}}

\caption{Optical conductivity obtained by the Kramers-Kronig transformation of
reflectance spectra for the (001) plane, {\bf E} $\parallel$ {\bf b} as a
function of temperature. Only TO $A_{u}$ modes should be seen in this
configuration. Inset: the "extra" mode $\sim$ 690 \cm.}

\label{FigSb}
\end{figure}

The curves of the {\bf b}-axis optical conductivity $\sigma_{b}(\omega) = \omega\,\mbox{Im}\,
\epsilon_{b}(\omega)/4\pi$, obtained by the
KK transform of the reflectivity spectra for each temperature, are
shown at Fig.~\ref{FigSb}. The conductivity is very illustrative to show a
remarkable difference between the $A_{u}^{1}$ and $A_{u}^{2}$ narrow modes and
the $A_{u}^{3}$ mode exhibiting a puzzling temperature transformation. The
sharpness of the $A_{u}^{1}$ and $A_{u}^{2}$ modes and the absence of the $B_{u}$
modes contribution to this spectrum confirms a good sample quality and its proper
orientation. Some deformations of the lineshapes for $T$ = 100 K and 7 K (in
particular, a small negative value of conductivity just below the mode
frequencies) are most likely the results of uncertainties of the KK
method, which is very sensitive to experimental errors in the regions where the
reflectivity approaches 0 or 1. We shall not attribute any physical meaning to
this effect.

\subsection{{\bf E} $\parallel$ {\bf ac}}

\begin{figure}[t]
\centerline{\epsffile{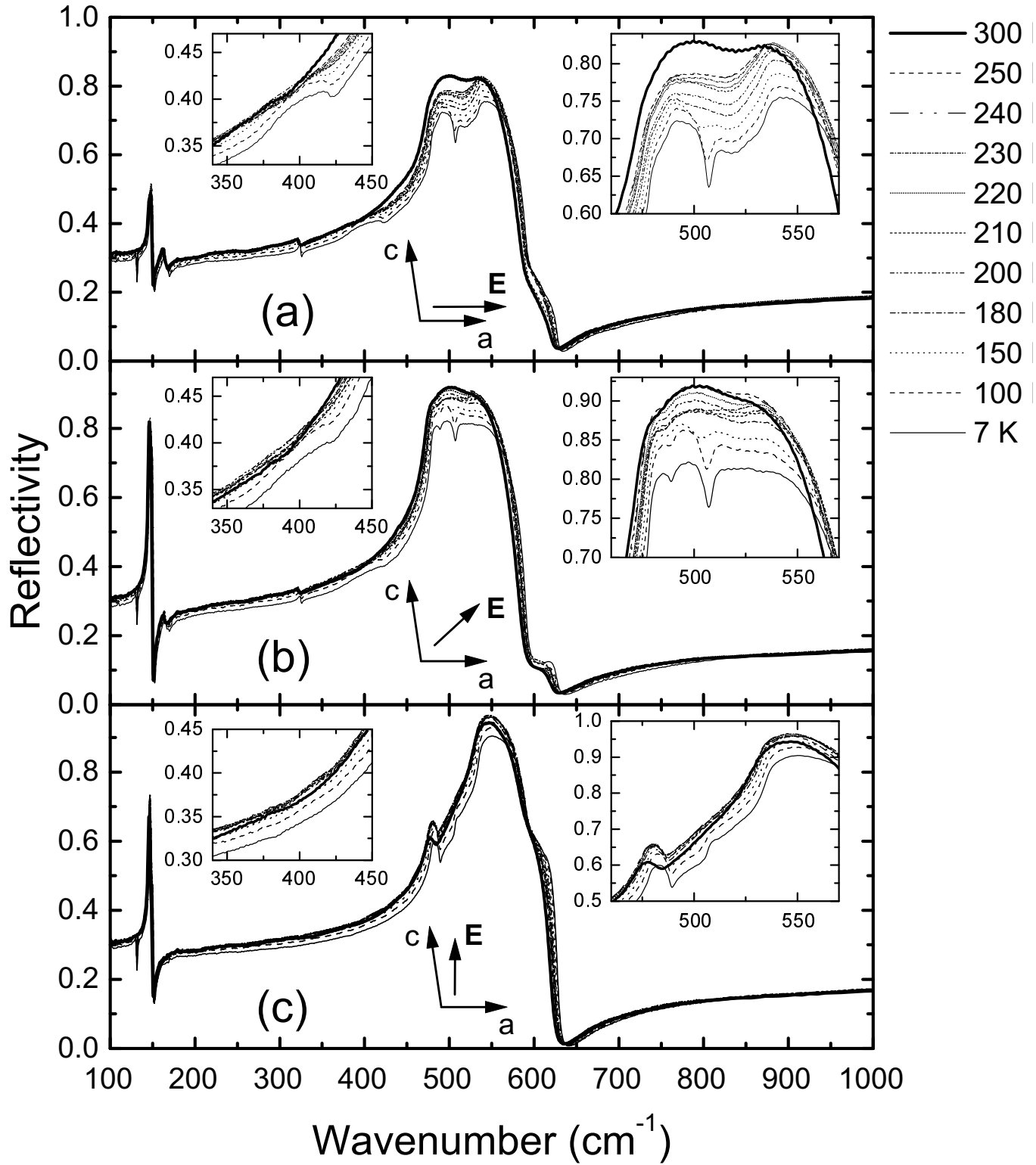}}

\caption{Reflectance spectra for the (010) plane for different polarizations
as a function of temperature. (a) {\bf E} $\parallel$ {\bf a} ($R_{00}$); (b)
$\widehat{({\bf E},{\bf a})}$=45\degree ($R_{45}$); (c){\bf E} $\perp$ {\bf a},
($R_{90}$).}

\label{FigRac}
\end{figure}

The reflectance spectra from the (010) plane for three polarizations of
the incident light ($R_{00}$, $R_{45}$ and $R_{90}$) are shown at
Fig.~\ref{FigRac}. The $B_{u}$ TO modes are expected to appear in these spectra.
The narrow mode $B_{u}^{1}$ ($\sim$ 145 \cm) is clearly observed in all
polarizations. Its intensity depends, of course, on the light polarization, i.e.
the angle between the electric field vector and the mode dipole
moment. Strong reststrahlen bands are seen in the 450 - 600 \cm\ range. The shape
and the center position of the band is polarization-dependent, which is
consistent with a suggestion that it is actually formed by at least two high
frequency intense modes. In the $R_{00}$ ({\bf E} $\parallel$ {\bf a}) and
$R_{45}$ spectra one can observe some minor contribution from the $A_{u}^{1}$
(162 \cm) and $A_{u}^{2}$ (323 \cm) modes. The possible reason is some
misorientation of the sample. A spike at about 130 \cm\ is of apparatus origin and should be
ignored.

\begin{figure}[t]
\centerline{\epsffile{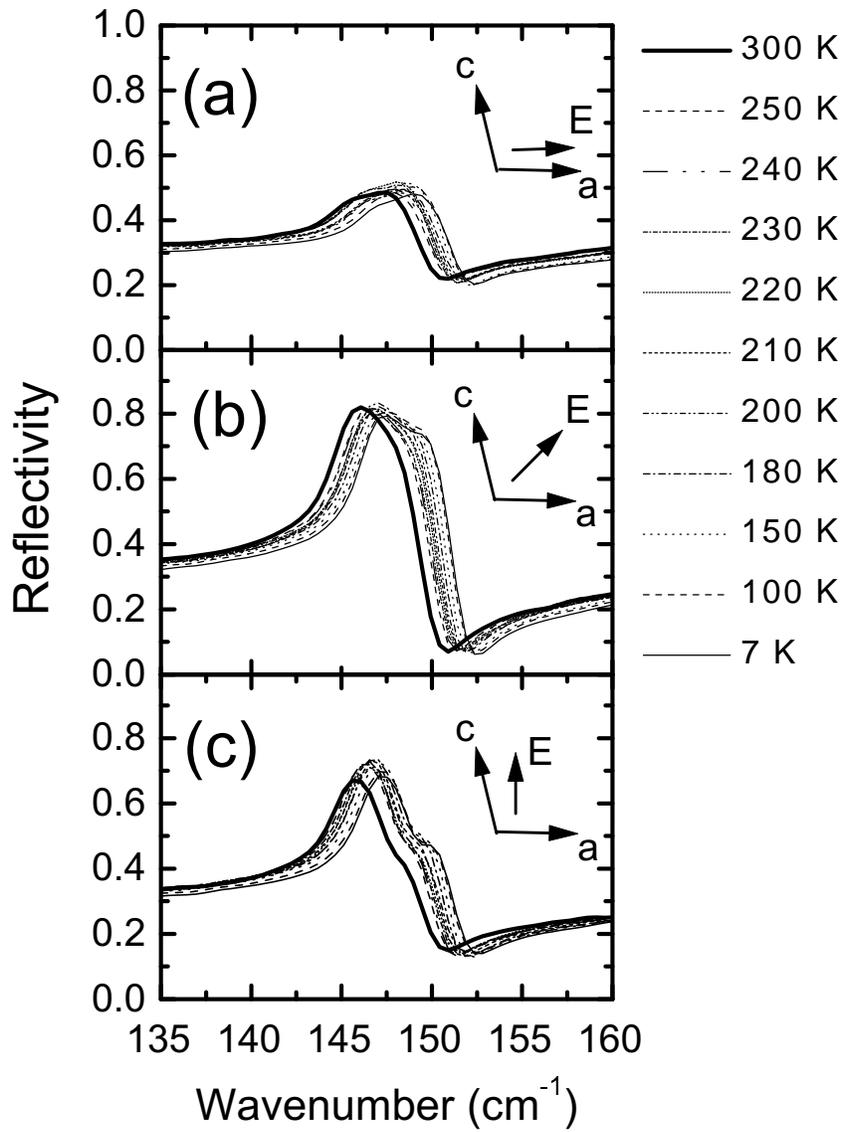}}

\caption{Reflectance spectra for the (010) plane for different polarizations
as a function of temperature (enlarged view of the $B_{u}^{1}$ mode region.
(a) {\bf E} $\parallel$ {\bf a} ($R_{00}$); (b) $\widehat{({\bf E},{\bf a})}$=45\degree ($R_{45}$);
(c){\bf E} $\perp$ {\bf a} ($R_{90}$).}

\label{FigMode145}
\end{figure}

As is in the case {\bf E} $\parallel$ {\bf b}, some extra structures are seen.
The first is a broad band in the range 380 - 425 \cm. It is especially pronounced for {\bf
E} $\parallel$ {\bf a}, less evident for intermediate polarization and absent for
{\bf E} $\perp$ {\bf a} (see the left insets in Fig.~\ref{FigRac}). A dip at 425
\cm, indoubtedly correlates with the dip at the same frequency at reflectance for
{\bf E} $\parallel$ {\bf b}. The second is a pronounced structure consisting of a peak at
$\sim$ 480 \cm\ and a dip at $\sim$ 485 \cm\ in the $R_{90}$ spectrum, existing
at all temperatures. In spite of proximity of this frequency to additional structure at
Fig.~\ref{FigRb}, a completely different temperature dependence makes one to separate these
modes. The third structure is a dip at $\sim$ 507 \cm\ on the top of the reststrahlen band
of $R_{00}$ and $R_{45}$ (see the right insets in Fig.~\ref{FigRac}). This frequency is very
close to the LO frequency of the $A_{u}^{3}$ mode, therefore it is most likely the $A_{u}^{3}$
mode which is seen in this spectrum for the same reason as the $A_{u}^{1}$ and $A_{u}^{2}$
modes are. The fourth structure is seen in the vicinity of the $B_{u}^{1}$ mode
(Fig.\,\ref{FigMode145}). The shape of this mode is such that it
is worth to suggest, that it is actually composed of two different modes (see,
especially, Fig.\,\ref{FigMode145}(c).

For each temperature a fitting procedure with introduction of 3 oscillators, corresponding to
principal $B_{u}$ modes has been performed. The phonon polarization angles were adjusted along
with other phonon parameters. Spectra $R_{00}$, $R_{45}$ and $R_{90}$ were fitted at the same time.
In the spirit of the FG-analysis prediciton, three lattice modes were introduced
for spectra fitting: one low-frequency, and two high-frequency modes. The fit
quality for T = 100 K is seen on the Fig.~\ref{FigRfit}(b)-(d). One can see that
the $B_{u}^{1}$ 145 \cm\ mode as well as the general shape of the reststrahlen band at
450 - 600 \cm\ are satisfactorily reproduced, confirming a suggestion that only 3
strong phonon modes are present. A bump at about 620 \cm\ is fitted without
invoking of additional Lorentzians: it results from the non-collinearity of the
mode and the incident radiation polarizations.

\begin{figure}[t]
\centerline{\epsffile{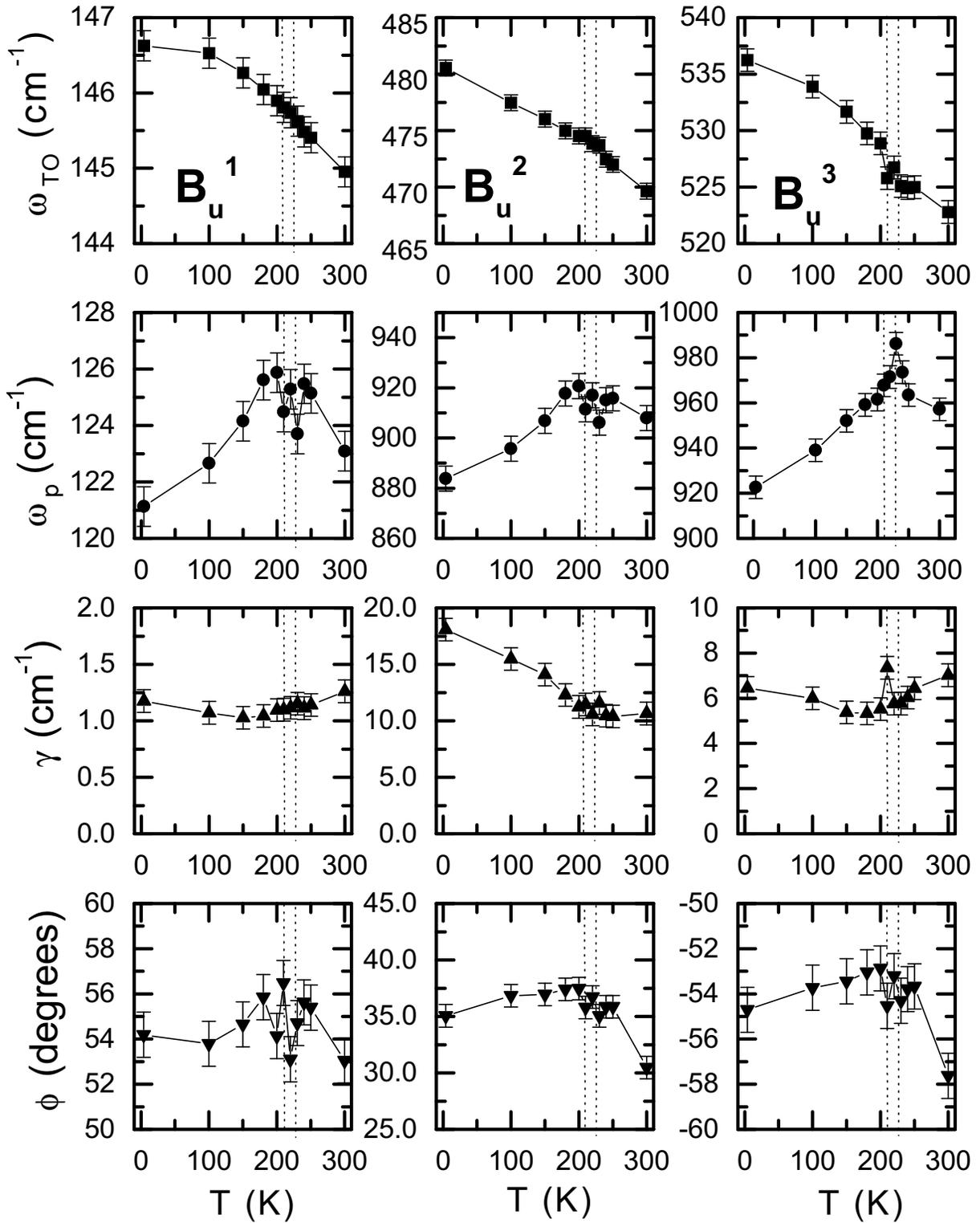}}

\caption{Parameters of the TO $B_{u}$ modes, obtained by the 3-mode dispersion analysis
of the reflectance spectra for the (010)-plane.
For each temperature three spectra for different polarizations of the incident light were fitted
simultaneously. Vertical dotted lines denote $T_{N1}$ and $T_{N2}$.}

\label{FigButo}
\end{figure}

The temperature dependence of the $B_{u}$ phonon parameters is presented at
Fig.~\ref{FigButo}. Unlike the case of the $A_{u}$ modes, there is no significant
difference in the temperature dependence of parameters of the low- and high-frequency
$B_{u}$ modes. All modes are monotonically hardening with cooling down, with some
positive kink at $T_{N}$ for the $B_{u}^{3}$ mode and negative kink for the
$B_{u}^{2}$ mode. The plasma frequencies of all modes have slight maximum at the
transition temperature. Two high frequency modes are much more intense than
the $B_{u}^{1}$ mode. The linewidth of all the modes
doesn't decrease with cooling down. Instead, it increases at low temperatures.
One should be careful in a straightforward interpretation of this result, because
the lineshape is not described perfectly, especially one of the high-frequency
modes. The true linewidth is better seen from the optical conductivity graph (see
below). The oscillator polarization angles do not significantly change with
frequency. The maximum angle change is 4 - 5\degree, while the change of the relative
angle between different oscillators polarization is less than 2 \degree. One can
state that within experimental errors oscillators angles almost doesn't change.

\begin{figure}[t]
\centerline{\epsffile{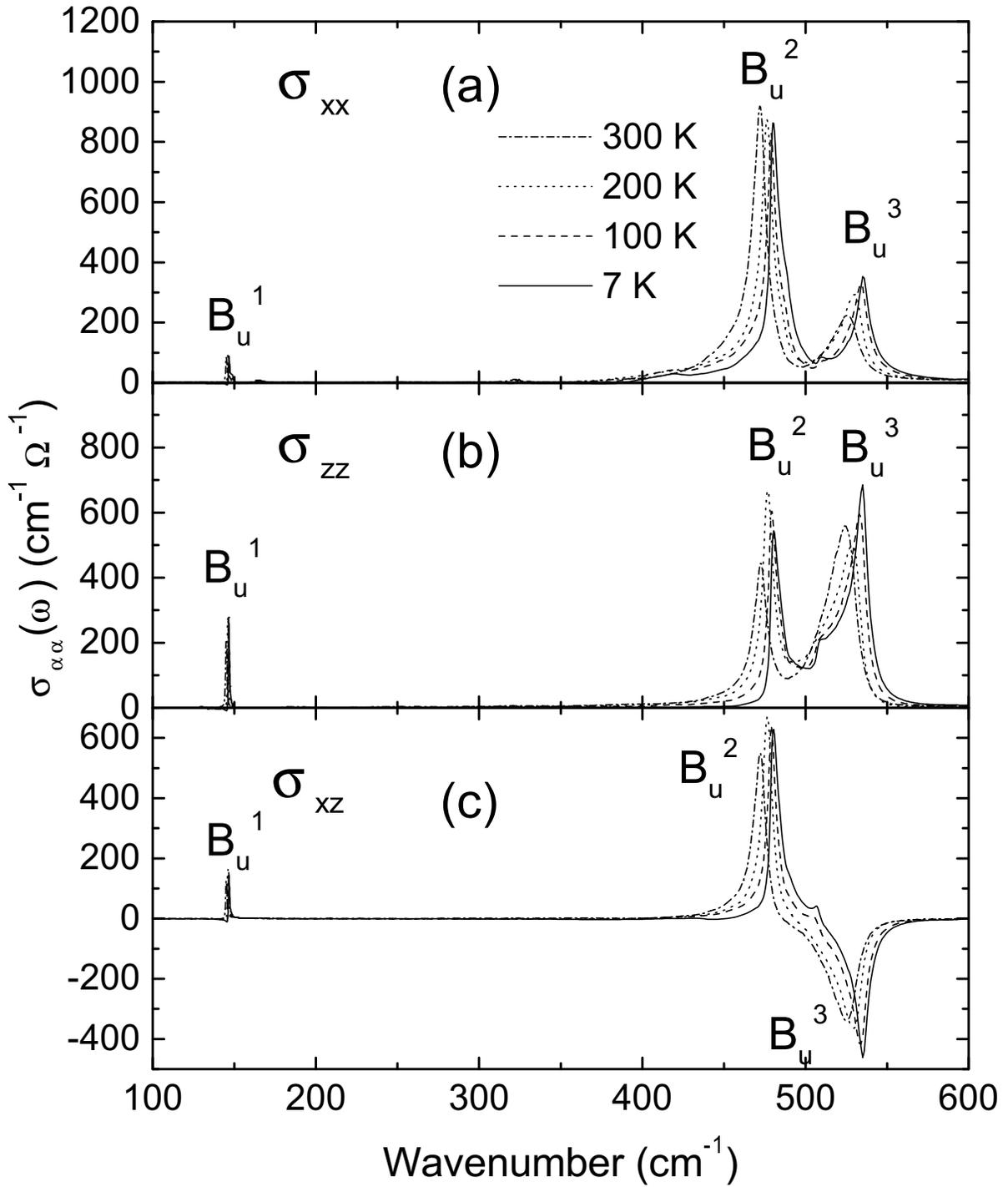}}

\caption{Components of the optical conductivity tensor
$\hat{\sigma}_{ac}(\omega)$ for selected temperatures. (a) $\sigma_{xx}$, (b)
$\sigma_{zz}$, (c) $\sigma_{xz}$.}

\label{FigSac}
\end{figure}

To investigate the true shape of the principal phonon modes and additional
structures, the KK analysis of the {\bf ac}-plane data for the
mentioned set of temperatures was implemented in the extended form
\cite{Kuzmenko2}. At Fig.~\ref{FigSac} all the components of the optical
conductivity tensor $\hat{\sigma}_{ac}(\omega) = \omega\,\mbox{Im}\,
\hat{\epsilon}_{ac}(\omega)/4\pi$ are plotted for selected temperatures.
Note that the off-diagonal component $\sigma_{xz}$ may have any sign unlike the diagonal
components $\sigma_{xx}$ and $\sigma_{zz}$ which must be positive.

The polarization angle of the $B_{u}^{2}$ mode is 35\degree, which is close to
the direction of the {[}101{]} chains. The $B_{u}^{3}$ mode is almost orthogonal
to the $B_{u}^{2}$ mode: the angle is -55\degree, which is close to the
{[}10$\overline{1}${]} chains direction. Therefore, with some approximation one
can state, that the $B_{u}^{2}$ and $B_{u}^{3}$ modes are stretches of the
{[}101{]} and {[}10$\overline{1}${]} chains correspondingly, which is in agreement
with several lattice dynamical calculations\cite{Kliche,Guha,Narang,Reichardt,Irwin1}. Note, that such
orthogonality is not determined by the crystal symmetry. The polarization angles of both
modes are almost temperature-independent.


\section{Discussion}

\subsection{Comparison with previous results}

A comparison of our data with previosly reported results of IR studies of CuO
cannot be direct, because we measured spectra, where the $A_{u}$ and $B_{u}$ modes
are separated and excited in purely transverse regime. All the quantitative and even
qualitative deviations with previous data (see below) can be ascribed to a
different way of spectra measurement and analysis.

At the Table \ref{TabFreqComp} the phonon frequencies at room temperature
previously obtained by means of infrared spectroscopy
\cite{Kliche,Guha,Narang,Homes} as well as neutron scattering \cite{Reichardt}
are collected. It is seen that the most serious discrepancy between reported
values of phonon frequencies takes place for modes $A_{u}^{3}$, $B_{u}^{2}$ and
$B_{u}^{3}$, which are very intense and manifest the largest LO-TO frequency
splitting. Pure TO mode should have the lowest possible frequency, which is in
nice agreement with the current result: our reported frequencies are smaller than
those reported in other IR spectroscopy papers. One should mention a
much better agreement between our data and the results of neutron scattering
experiments\cite{Reichardt}, where characteristics of pure TO modes were
determined as well.

The softening of the $A_{u}^{3}$ mode was reported by Homes {\it et al}
\cite{Homes}, who observed a sudden drop of the phonon frequency at the N\'eel
transition from 450 to 430 \cm, i.e. $\sim$ 5 \%. Our data qualitatively confirm
this interesting result. We observe even stronger effect: the TO frequency drops
from 410 to 370 \cm\ ($\sim$ 10 \%), which is twice as large as was reported
(Fig.~\ref{FigAuto}).

Ref.~\cite{Homes}, was thus far the only paper, to our knowledge, where IR
spectra of single-crystal CuO at low temperatures were studied. Authors didn't
report any new IR-active modes below the N\'eel temperature; it was implied, that no additional lines
are present at higher temperatures either. However, an absence
of extra IR-active modes is quite strange, if one compares it with an observation of at
least 5 "unexpected" lines in the Raman spectra\cite{Chen}. Conversely, according to our
data, several "extra" modes are present in IR spectra in the whole 7 - 300 K frequency
range. We believe, that better orientation of the wave vector and polarization of the incident
radiation may facilitate observation and analysis of minor IR-active modes.

\subsection{The $A_{u}^{3}$ mode anomaly}

\begin{figure}[t]
\centerline{\epsffile{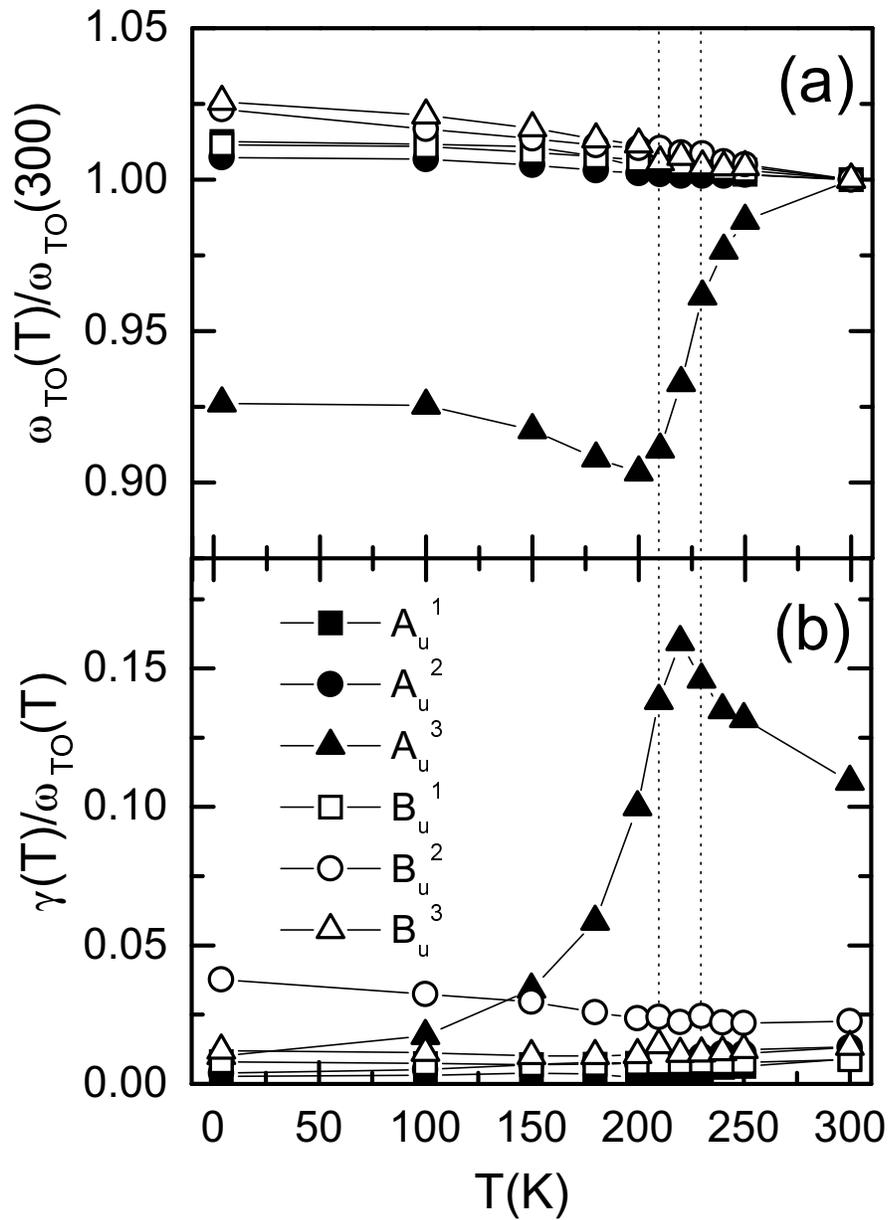}}

\caption{Relative characteristics of the principal IR-active phonon modes in CuO.
(a) The RT-normalized frequency; (b) the relative linewidth. Vertical dotted
lines denote $T_{N1}$ and $T_{N2}$.}

\label{FigRelPar}
\end{figure}

The $A_{u}^{3}$ mode among other principal IR-active modes behaves in the most
anomalous way. It is demonstrated at Fig.~\ref{FigRelPar}, where relative
RT-normalized TO frequencies and relative linewidths of all 6 principal modes are plotted
together on the same graph. A close relation between magnetic ordering transition
at 213 - 230 K and temperature transformations of this mode is without any doubt.

At 300 K and, especially, at the transition temperature the mode is anomalously
broad (12 - 15 \%), which indicates, that it is strongly coupled to some system
quasiparticles. The most probable candidates are the low-energy magnetic
excitations. If one propose, that there is a strong coupling between spin
excitations and the $A_{u}^{3}$ lattice mode, the temperature transformations of
the mode can be explained by reconstruction of the magnon spectrum upon cooling
down below the N\'eel temperature. It is well
established\cite{Ain,OKeeffe,Chattopadhyay}, that spin correlations in the
AFM {[}10$\overline{1}${]} chains are present well above the N\'eel temperature.
At high temperatures one can consider the magnetic interaction to be of quasi 1D
character. In the 1D Heisenberg AFM chains with S=1/2 and exchange $J$ there is a
continuum of triplet excitations with lower and upper boundary
curves\cite{Cloizeaux,Yamada} $\epsilon_{\mbox{\scriptsize min}}(q)=(\pi J/2) \sin(q)$ and
$\epsilon_{\mbox{\scriptsize max}}(q)=\pi J \sin(q/2)$. A large linewidth of the $A_{u}^{3}$ mode
can be explained by its interaction with continuum of magnetic excitations. Below
the N\'eel point an exchange interaction in another directions gives rise to
long-range magnetic ordering and a continuum of spin excitations should collapse
to the magnon dispersion curves, which, in turn, results in narrowing of the
phonon mode.

The reason for an exceptionally strong interaction of the $A_{u}^{3}$ mode with
spin excitations one should look for in analysis of its eigenvector. Several
lattice dynamical calculations were performed
\cite{Kliche,Guha,Narang,Reichardt,Irwin1} which yielded the lattice mode
eigenvectors. According to results of all the calculations this mode is
characterized by the largest displacement of the oxygen atoms along the {\bf
b}-axis. It results also in a large dipole moment of this mode. As a consequence,
the Cu-O-Cu chain angle experiences the largest variation, when the $A_{u}^{3}$
mode is excited. The copper spins are coupled via the superexchange interaction,
which is very sensitive to the Cu-O-Cu angle. In the {[}10$\overline{1}${]}
chains the angle is equal to 146\degree\ (Fig.~\ref{FigChains}), which is close
to the 180\degree\ superexchange. It gives a negative exchange constant and AFM
interaction. However, the 90\degree\ superexchange is expected to be
positive\cite{Goodenough} (an indirect confirmation is the ferromagnetic exchange
along the {[}101{]} chain with angle equal to 109\degree), and there exists some
intermediate angle, where the superexchange is changing sign. Therefore, the
motion of atoms, corresponding to excitation of the $A_{u}^{3}$ mode can
significantly vary the value of the superexchange coupling constant. A strong chain
bending probably could alternate the exchange sign.

Due to energy and momentum conservation a zone-center phonon can couple to a pair
of magnons (bi-magnon) having opposite wave-vectors and one-half frequency of the
phonon. Another option is interaction with a single zone-center optical magnon of
the same energy. The magnon dispersion curves has been studied by inelastic
neutron scattering~\cite{Ain,Chattopadhyay}. One acoustic and one optical branch
were observed; the energy of optical magnon at the $\Gamma$ point is 5.6 THz (187
\cm), which is very close to one-half of the $A_{u}^{3}$ mode frequency (370 - 380
\cm\ at low temperatures), while no optical magnons near 370 - 410 \cm\ were observed.
Therefore, the "bi-magnon" or, in particular "optical bi-magnon" scenario of
resonant phonon-magnon coupling is the most probable. A more detailed theory of
this effect has to be fabricated.

\subsection{Other signatures of spin-phonon interaction}

The anomalous properties of the $A_{u}^{3}$ mode is not the only
manifestation of the spin-phonon interaction in CuO (although the
most prominent one). Other modes also demonstrate some anomalies,
most likely related to the magnetic ordering.

One type of anomalies is unusually strong hardening of some "extra" modes below
$T_{N}$. The most outstanding example is reported in this work hardening of the 690 \cm\
mode from 650 \cm\ at 150 K to 690 \cm\ at 7 K and strong hardening
of the the Raman-active 240 \cm\ mode found by Chen {\it et al} \cite{Chen}.
To our mind, these effects are related. We would follow the idea of explanation given in
Ref.~\cite{Chen}. The phenomenon can be viewed as a non-resonant spin-phonon interaction,
when spin products in the Heisenberg Hamiltonian can be approximated by their effective averages.
In this case the temperature dependence of the phonon frequency is expressed by:

\begin{equation}
\omega^{2}_{n}(T)=\omega^{2}_{n0}+\sum_{ij}C_{ij}^{n}\cdot\langle{\bf S}_{i}{\bf S}_{j}\rangle
(T)\,.
\end{equation}

\noindent The $C_{ij}^{n}$ coefficients are characteristics of spin-phonon interaction; in
principle, they may have any sign, depending on the eigenvector of the particular
phonon mode. The mode strong hardening is supposedly due to the second "spin"
term: magnetic ordering gives additional lattice rigidity.

The same phenomenon is possibly responsible for another anomaly, namely,
the change of slope $\partial\omega_{\mbox{\scriptsize TO}}/\partial T$ of the principal phonon
modes at the N\'eel point (see Figs.~\ref{FigAuto} and \ref{FigButo}). It is seen that for the
modes $A_{u}^{1}$, $A_{u}^{2}$ and $B_{u}^{3}$ the slope is higher in the AFM
phase, while for the $B_{u}^{2}$ mode the slope is higher at $T>T_{N}$; for the
$B_{u}^{1}$ mode it doesn't change at all. Such a differences can be explained by
different values and signs of coefficients $C_{ij}^{n}$.

\subsection{"Extra" zone-center modes and zone folding}

Activation of additional phonon modes in infrared and Raman spectra is
usually a signature of unit cell multiplication and zone folding.
Such activation is in effect in CuO. As is stated above, several "extra" IR-active modes
are observed. In the Raman spectra 5 new modes
were detected at low temperatures \cite{Chen}. Authors related their
frequencies to the phonon dispersion curves obtained by
inelastic neutron scattering \cite{Reichardt} at the zone boundary point.
In Ref.\,\cite{Chen} this point was referred to as $Z'$. We follow the original
notation and designate it by X (see Fig.\,4 in Ref.\,\cite{Reichardt}).

\begin{figure}[t]
\centerline{\epsffile{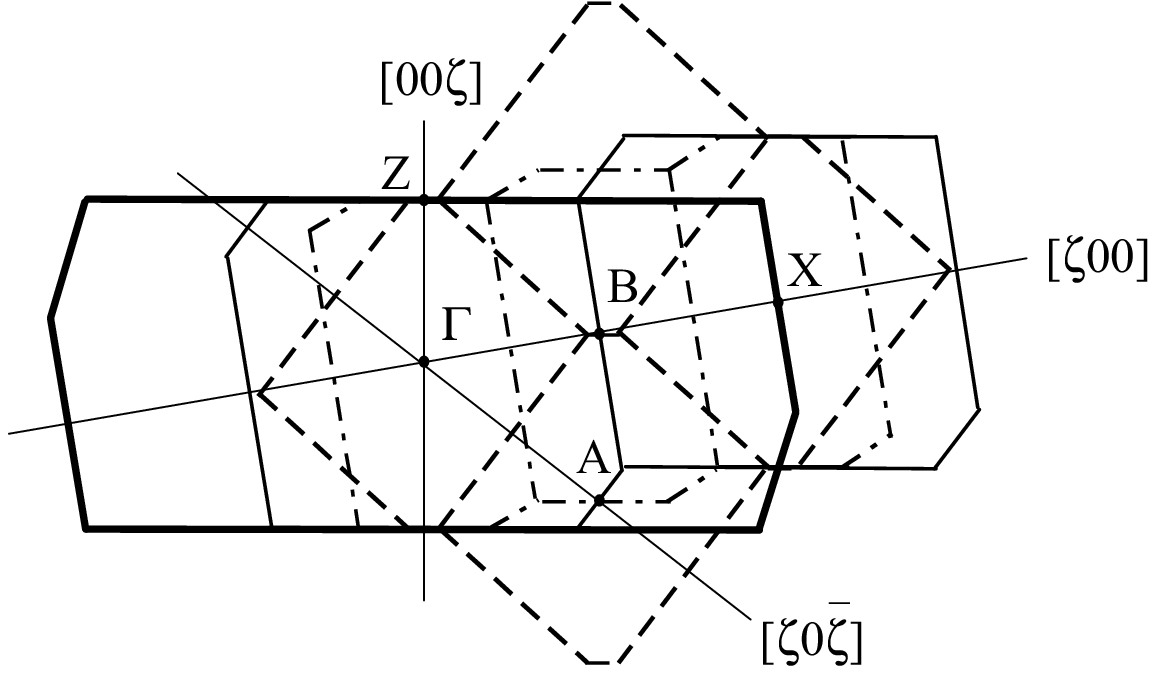}}

\caption{The {\bf ac}-plane projection of the Brillouin zones corresponding to
different unit cells. Thick solid line - primitive cell \{{\bf a}, ({\bf a}+{\bf
b})/2, {\bf c}\}, solid line - unit cell \{{\bf a}, {\bf b}, {\bf c}\}, dashed
line - \{{\bf a}+{\bf c}, {\bf b}, {\bf a}-{\bf c}\}, dashed-dotted line \{2{\bf
a}, {\bf b}, {\bf c}\}. }

\label{FigBz}
\end{figure}

At Fig.\,\ref{FigBz} the Brillouin zones corresponding to several
schemes of the unit cell multiplication are drawn (a projection to the {\bf ac}-plane).
In these schemes different symmetry points fold to the zone center. It is easy to prove, that folding
of the X-point to the zone center is equivalent to disappearance of the non-trivial
translation ({\bf a}+{\bf b})/2, or base-centering of the space-group. In other words,
activation of the phonons from the X point requires that the unit cell should become the
primitive one.

In the Table~\ref{TabExtra} "extra" IR active and Raman-active mode
frequencies are collected. Each mode frequency is related to close
phonon energy (if any) in symmetry points X(1, 0, 0), A(1/2, 0, -1/2),
B(1/2, 0, 0) or C(0, 1/2, 0) at 300 K. To make comparison more reliable,
the "extra" modes frequencies are taken at temperature as close as
possible to 300 K. One can see, that many modes (all Raman-active
and several IR-active ones) have close analogs at the X point. At the same
time, other IR-active modes (147 \cm, 165 \cm, 475 \cm, 480 \cm)
could be phonons from the A point, or, in some cases, from the B or C
points. The 690 \cm\ mode is a special
case: due to strong hardening of this mode with cooling and failure
to observe it at high temperatures, the comparison to the 300 K
dispersion data is impossible. In addition, the mode energy is higher than
the upper limit of the reported frequency region in the neutron scattering
experiments (20 THz, or 667 \cm). The phonon dispersion curves at low temperatures in broader
energy interval are desirable.

Although some "extra" modes have analogs in the B or C points, folding of the A and X points
to the zone center (the A + X folding scheme) is the simplest option to explain appearance of
all "extra" modes. It corresponds to the "diagonal" doubling of the unit cell
\{{\bf a}, {\bf b}, {\bf c}\} $\rightarrow$ \{{\bf a}+{\bf c}, {\bf b}, {\bf
a}-{\bf c}\} (see Fig.\,\ref{FigBz}). This scenario looks attractive,
because new crystal axes correspond to principal anisotropy directions for several physical
properties (exchange constants, sound velocity, principal dielectric axes etc.), and the
unit cell is the same as the magnetic unit cell in the AFM phase \cite{Yang}.

In the framework of this scenario the $A_{u}^{3}$ principal mode, the 485 \cm\ and the 425 \cm\
"extra" modes should be the phonon modes from the points $\Gamma$, A and X correspondingly
belonging to the same dispersion branch, according to "rigid-ion" modelling of dispersion curves
by Reichardt {\it et al} \cite{Reichardt}. It is in a good agreement with some similarity of
temperature dependences of parameters of these modes (see Figs. \ref{FigAuto} and \ref{FigAutoext}).

In summary, we propose, that in reality the crystal structure is more complicated, than was considered
before; it is the case, to our mind, already at room temperature, because most of "extra" IR-active
modes are present in the whole temperature range. On the base of the fact that IR-active and Raman-active
modes have different frequencies one can suggest that the crystal space-group is still
centro-symmetric, although the copper atoms are not necessarily located
in the $C_{i}$ position. However, the alternative space-group for the CuO, proposed by
Asbrink and Waskowska, is not centro-symmetric (C$_{s}^{4}$); the solution of this
mismatch is unclear.

One of the central issues is the mechanism of the unit cell multiplication and
formation of superstructures. There exist a variety of examples of such effect,
when "extra" IR and/or Raman modes are emerging, which are undoubtedly the
zone-boundary folded phonons. In the extensively studied compounds
CuGeO$_{3}$ and $\alpha'$-NaV$_{2}$O$_{5}$ the unit cell doubles as a result of
spin-phonon interaction, and new IR modes are observed \cite{Damascelli,Popova}.
Another example present compounds with a charge disproportionation, out of those the BaBiO$_{3}$ is
probably the most famous one. In this system the bismuth is disproportionated
according to scheme: 2\,Bi$^{4+}$\,$\rightarrow$\,Bi$^{3+}$\,+\,Bi$^{5+}$, and atoms
in different valence states form superstructure, which is a reason for
"extra" quite strong IR line\cite{Uchida}. One cannot exclude a collective Jahn-Teller
effect as an engine of superlattice formation. The question about particular type
of spin-charge-lattice ordering in CuO is open. It could be closely connected
with formation of inhomogeneity phases in cuprates.

\subsection{High-frequency dielectric function}

The reflectivity and the dielectric function in the mid-infrared range well above
the maximum phonon energy ($\sim$ 0.08 eV) but below the optical gap ($\sim$ 1.3
eV) are determined by the electronic polarizability. We observe significant
difference between values of mid-infrared reflectivity for different
polarizations. It results in an appreciable anisotropy of the high-frequency
dielectric tensor. The temperature dependence of the mid-IR reflectivity is
small, therefore, we discuss only room-temperature data. All the components of
the $\hat{\epsilon}^{\infty}$ can be found from the dispersion analysis of
spectra, see formulas (\ref{epsb}), (\ref{epsac}).

The smallest value of $\epsilon^{\infty}$ is observed along the {\bf b}-axis:
$\epsilon^{\infty}_{b} = 5.9$; it is apparently one of the prinicipal values of
the dielectric tensor. The diagonalization of the tensor gives directions and
values of maximal and minimal levels of $\epsilon^{\infty}$ within the ({\bf
ac})-plane. The maximal value $\epsilon^{\infty}_{\mbox{\scriptsize max}} = 7.8$
is observed along the direction, corresponding to the angle
$\phi_{\mbox{\scriptsize max}}=-36^{\circ}$, the minimal value
$\epsilon^{\infty}_{\mbox{\scriptsize min}} = 6.2$ is along the orthogonal
direction for $\phi_{\mbox{\scriptsize min}}=54^{\circ}$. One can see that
direction $\phi_{\mbox{\scriptsize max}}$ is very close to direction of the
{[}10$\overline{1}${]} chains ($\phi_{\mbox{\scriptsize {[}10$\overline{1}${]}}}
=  -42.5^{\circ}$) while $\phi_{\mbox{\scriptsize min}}$ almost exactly
corresponds to $\phi_{\mbox{\scriptsize {[}101{]}}} = 52.8^{\circ}$. In other
words, the high-frequency dielectric constant within the ({\bf ac})-plane is
maximal along the {[}10$\overline{1}${]} direction, and minimal along the
{[}101{]} direction. Authors \cite{Ito} have measured the high-frequency
dielectric function of polycrystalline sample by the ellipsometric method. Their
reported value $\epsilon^{\infty} = 6.45$ is in a good agreement with the average
quantity
$(\epsilon^{\infty}_{xx}+\epsilon^{\infty}_{yy}+\epsilon^{\infty}_{zz})/3 = 6.6$
obtained here.

\subsection{Effective charges}

Another important value derived from infrared spectra is an effective ionic
charge. There are several definitions of the effective charge. We would mention
the Born, or "transverse" charge $e^{\ast}_{\mbox{\scriptsize T}}$, the Szigeti charge
$e^{\ast}_{\mbox{\scriptsize s}}$ and the Scott charge $Ze$ \cite{Gervais}. The "transverse"
charges can be calculated using the sum-rule:

\begin{equation}
\sum_{i}\omega^{2}_{\mbox{\scriptsize p},i} =
\frac{4\pi}{v_{c}}\,\sum_{k}\frac{(e^{\ast}_{\mbox{\scriptsize T} ,k})^{2}}{m_{k}}\,,
\end{equation}

\noindent where $v_{c}$ is the primitive cell volume, the sum in the left side is
over IR-active modes, the sum in the right side is over all atoms in the
primitive cell. For the binary compound this relation in combination with the
electric neutrality condition directly yields the "transverse" charge of both
atoms. In CuO due to anisotropy the transverse charge slightly differs for
different directions. The values along the [010] ({\bf b}-axis) the ${[}10{1]}$
and the ${[}10\overline{1}{]}$ directions at room temperature are 1.92, 2.04 and
2.10 correspondingly (the $\omega_{\mbox{\scriptsize p}, i}$ are taken from the Figs.\,\ref{FigAuto},
\ref{FigButo}).

The Scott charge and Szigeti charges are related to the "transverse" charge:

\begin{equation}
Ze = \frac{e^{\ast}_{\mbox{\scriptsize T}}}{\sqrt{\epsilon^{\infty}}}\,\,\,,\,\,\,e^{\ast}_{s} =
\frac{3\,e^{\ast}_{\mbox{\scriptsize T}}}{\epsilon^{\infty}+2}\,.
\label{charges}
\end{equation}

\noindent The value of $Ze$ divided to the nominal valence is often considered as
the degree of ionicity \cite{Gervais}. For the CuO it appears to be about 40 \%.
The Szigeti charge is useful in the context of the polarized point ions model \cite{Szigeti}.
In literature all these charges are used, so that it is reasonable to report all of them.

In the Table \ref{TabEps} oxygen effective charges for the CuO along with other
related copper oxides (Cu$_{2}$O, La$_{2}$CuO$_{4}$, Nd$_{2}$CuO$_{4}$ and
YBa$_{2}$Cu$_{3}$O$_{6}$) are collected. It is seen that charge values in CuO
reasonably agree to the corresponding values in other relevant oxides.

\section{Conclusions}

We have measured far- and mid-infrared reflectivity spectra of monoclinic CuO
from the (010) and (001) crystal faces in wide temperature range. We obtained for
the first time characteristics of pure TO $A_{u}$ and $B_{u}$ modes separately.
Our data finally confirm that there are 3$A_{u}$ + 3$B_{u}$ intense modes, in
accordance with prediction of the FG-analysis for the $C_{2h}^{6}$ space-group.

We report existence of several "extra" less intense IR-active
modes in CuO. Analysis of the phonon dispersion curves leads to the conclusion that each "extra"
IR-active as well as reported earlier \cite{Chen} Raman-active mode could be folded
phonon from either X (1, 0, 0) or A (1/2, 0, 1/2) symmetry points.
Such folding is compatible with the "diagonal" doubling of the unit cell with the new basis
\{{\bf a}+{\bf c}, {\bf b}, {\bf a}-{\bf c}\}. So the space-group in reality is
lower than that was considered, but still centro-symmetric.

The 690 \cm\ "extra" IR-active mode exhibits anomalous hardening, similar to behaviour of
the 240 \cm\ Raman mode; the reason could be in additional rigidity of lattice
due to magnetization, a special manifestation of the spin-phonon interaction.
Another effect, which can be explained in a similar way, is a slope change of the
phonon frequencies vs. temperature at the N\'eel point.

The anomalous softening and narrowing of the $A_{u}^{3}$ mode 410 \cm\ we explain
by its strong resonance coupling to the optical or acoustic bi-magnons. Reconstruction of magnetic
excitations spectrum at the AFM transition strongly affects the phonon characteristics.

In summary, the CuO demonstrates a variety of anomalous properties, which show
complex interplay of spin, charge, and phonon subsystems already in the simplest
copper(II) oxide. A further insight into physics of CuO may contribute to
elaboration of non-contradictory picture of antiferromagnetism and
superconductivity in the high-$T_{c}$ cuprates.

\acknowledgements

This investigation was supported by the Netherlands Foundation
for Fundamental Research on Matter (FOM) with financial aid from
the Netherlandse Organisatie voor Wetenschappelijk Onderzoek
(NWO). The activity of A.B.K., E.A.T. and A.A.B. was also supported
by the Russian Foundation for Basic Research (RFBR), grant No
99-02-17752. A lot of thanks we address to H. Bron and F. van der Horst (University of
Groningen) for invaluable help in samples characterization.

\newpage

\begin{table}[h]
\caption{Frequencies of the principal $A_{u}$ and $B_{u}$ modes (in \cm) at 300 K obtained by
different research groups.}
\begin{tabular}{ccccccc}
  Mode & Kliche& Guha & Narang\tablenotemark[1] & Homes & Reichardt\tablenotemark[2] & This \\
       & {\it et al} \cite{Kliche} & {\it et al} \cite{Guha} & {\it et al} \cite{Narang} & {\it et al} \cite{Homes} & {\it et al} \cite{Reichardt} & work \\
  \hline
  $A_{u}^{1}$ & 161 & 168 & 163 & 166.5 & 160 & 160.5 \\
  $A_{u}^{2}$ & 321 & 325 & 324 & 323.7 & 326 & 321.5 \\
  $A_{u}^{3}$ & 478 & 465 & 444 & 450.0 & 423 & 409.0 \\
  $B_{u}^{1}$ & 147 & 150 & 147 & 147.6 & 142 & 144.9 \\
  $B_{u}^{2}$ & 530 & 510 & 515 & 516.0 & 480 & 468.0 \\
  $B_{u}^{3}$ & 590 & 570 & 586 & 566.0 & 520 & 522.5 \\
\end{tabular}
\tablenotetext[1]{Data for the 77 K} \tablenotetext[2]{Neutron data}
\label{TabFreqComp}
\end{table}

\begin{table}
\caption{Frequencies (in \cm) of zone-center "extra" modes, obtained from IR and Raman
spectra, and close phonon frequencies in some off-center symmetry points of the
Brillouin zone, obtained by inelastic neutron scattering at 300 K. IR and Raman data
(with the exception of the 690 \cm\ mode) are presented at temperatures as close as possible
to 300 K, }

\begin{tabular}{lcccccc}
  \multicolumn{2}{c}{IR/Raman\cite{Chen} data} &\ & \multicolumn{4}{c}{Inelastic neutron scattering data \cite{Reichardt} }\\
  \cline{1-2} \cline{4-7}
  Acti- & Frequ-&\ &     X     &       A       &       B     &      C      \\
  vity  & ency  &\ & (1,0,0) & (0.5,0,0.5) & (0.5,0,0) & (0,0.5,0) \\
  \cline{1-2} \cline{4-7}
  IR    & 690 &\ & ?   & ?   & ?   & ?   \\
  IR    & 630 &\ & 630 & -   & 633 & -   \\
  Raman & 507 &\ & 513 & -   & -   & -   \\
  IR ({\bf E} $\parallel$ {\bf b}) & 480 &\ & -   & 485 & 473 & 487 \\
  IR ({\bf E} $\parallel$ {\bf ac})& 475 &\ & -   & 485 & 473 & 487 \\
  IR    & 430 &\ & 437 & -   & -   & -   \\
  Raman & 331 &\ & 330 & -   & -   & -   \\
  Raman & 240 &\ & 243 & -   & -   & -   \\
  Raman & 218 &\ & 220 & 223 & -   & -   \\
  Raman & 175 &\ & 177 & -   & -   & -   \\
  IR    & 165 &\ & -   & 167 & -   & -   \\
  IR    & 147 &\ & -   & 143 & -   & 150 \\
\end{tabular}

\label{TabExtra}
\end{table}

\begin{table}
\caption{A comparison of oxygen effective charges (in electrons) for the cupric
oxide CuO and relevant copper oxides. When authors presented only one type of
effective charge, other values are recalculated using $\epsilon^{\infty}$.}

\begin{tabular}{lcccc}
Compound & $\epsilon^{\infty}$ & $e^{\ast}_{s}$ & $Ze$ & $e^{\ast}_{T}$
\\ \hline

CuO (${\bf E} \parallel $ ${[}010{]}$)                            & 5.9 & 0.76  & 0.81
& 1.96 \\

CuO (${\bf E} \parallel $ ${[}101{]}$)                            & 6.2 & 0.75  & 0.82
& 2.06 \\

CuO (${\bf E} \parallel $ ${[}10\overline{1}{]}$)                 & 7.8 & 0.64  & 0.76
& 2.12 \\

Cu$_{2}$O \cite{Noguet}                                     & 6.5 & 0.68  & 0.75  & 1.91 \\

La$_{2}$CuO$_{4}$ (${\bf E} \parallel {\bf c}$) \cite{Gervais88}        & 4.1 & 1.0   & 1.0   & 2.0 \\

Nd$_{2}$CuO$_{4}$ (${\bf E} \parallel {\bf c}$) \cite{Abrosimov}        & 4.2 & 1.1   & 1.1   & 2.2 \\

YBa$_{2}$Cu$_{3}$O$_{6}$ (${\bf E} \parallel {\bf c}$) \cite{Bazhenov}  & 4.5 & 0.8  & 0.8   & 1.7

\end{tabular}
\label{TabEps}
\end{table}

\onecolumn


\newpage

\newpage


\newpage


\newpage


\newpage


\newpage


\newpage


\newpage


\newpage


\newpage


\newpage






\newpage






\newpage


\newpage


\end{document}